\documentclass{emulateapj}
\usepackage{apjfonts,rotating,eqnarray}
\usepackage{color}
\usepackage{amsmath}

\def\asca{{\itshape ASCA\/}}

\def\chandra{{\itshape Chandra\/}}

\def\hst{{\itshape HST\/}}

\def\spitzer{{\itshape Spitzer\/}}
\def\herschel{{\itshape Herschel\/}}
\def\galex{{\itshape GALEX\/}}

\def\xray{\hbox{X-ray}}

\def\etal{{et\,al.}}

\def\ltsima{$\; \buildrel < \over \sim \;$}
\def\simlt{\lower.5ex\hbox{\ltsima}}
\def\gtsima{$\; \buildrel > \over \sim \;$}
\def\simgt{\lower.5ex\hbox{\gtsima}}
\def\kms{\ifmmode{~{\rm km~s^{-1}}}\else{~km s$^{-1}$}\fi}
\def\lsim{\lower0.3em\hbox{$\,\buildrel <\over\sim\,$}}
\def\gsim{\lower0.3em\hbox{$\,\buildrel >\over\sim\,$}}

\def\msol{$M_\odot$}
\def\h2{H$_2$}
\def\flux{erg~cm$^{-2}$~s$^{-1}$}
\def\lum{erg~s$^{-1}$}

\def\arcsec{\mbox{$^{\prime\prime}$}}

\def\sfr{$M_{\odot}$~yr$^{-1}$}

\def\aap{A\&A}
\def\apj{ApJ}
\def\apjl{ApJL}
\def\apjs{ApJS}
\def\aj{AJ}
\def\mnras{MNRAS}
\def\araa{ARA\&A}

\begin{document}

\shortauthors{LEHMER ET AL.}
\shorttitle{X-ray Binary Formation in M51}

%
\title{On the Spatially Resolved Star-Formation History in M51 II: X-ray Binary Population Evolution}
%

\author{
B.~D.~Lehmer,\altaffilmark{1}
R.~T.~Eufrasio,\altaffilmark{1}
L.~Markwardt,\altaffilmark{1}
A.~Zezas,\altaffilmark{2,3,4}
A.~Basu-Zych,\altaffilmark{5,6}
T.~Fragos,\altaffilmark{7}
A.~E.~Hornschemeier,\altaffilmark{6} 
A.~Ptak,\altaffilmark{6} 
P.~Tzanavaris,\altaffilmark{6}
\& M.~Yukita\altaffilmark{8,6}
}

\altaffiltext{1}{Department of Physics, University of Arkansas, 226 Physics
Building, 825 West Dickson Street, Fayetteville, AR 72701, USA}
\altaffiltext{2}{University of Crete, Physics Department \& Institute of Theoretical \& Computational Physics, 71003 Heraklion, Crete, Greece}
\altaffiltext{3}{Foundation for Research and Technology-Hellas, 71110 Heraklion, Crete, Greece}
\altaffiltext{4}{Harvard-Smithsonian Center for Astrophysics, 60 Garden Street,
Cambridge, MA 02138, USA}
\altaffiltext{5}{Center for Space Science and Technology, University of
Maryland Baltimore County, 1000 Hilltop Circle, Baltimore, MD 21250, USA}
\altaffiltext{6}{NASA Goddard Space Flight Center, Code 662, Greenbelt, MD
20771, USA} 
\altaffiltext{7}{Geneva Observatory, Geneva University, Chemin des Maillettes
51, 1290 Sauverny, Switzerland}
\altaffiltext{8}{The Johns Hopkins University, Homewood Campus, Baltimore, MD
21218, USA}

%
\begin{abstract}
%

We present a new technique for empirically calibrating how the \xray\
luminosity function (XLF) of \xray\ binary (XRB) populations evolves following
a star-formation event.  We first utilize detailed stellar population synthesis
modeling of far-UV to far-IR photometry of the nearby face-on spiral galaxy M51
to construct maps of the star-formation histories (SFHs) on subgalactic
($\approx$400~pc) scales.  Next, we use the $\approx$850~ks cumulative
\chandra\ exposure of M51 to identify and isolate 2--7~keV detected point
sources within the galaxy, and we use our SFH maps to recover the local
properties of the stellar populations in which each \xray\ source is located.
We then divide the galaxy into various subregions based on their SFH properties
(e.g., star-formation rate [SFR] per stellar mass [$M_\star$] and mass-weighted
stellar age) and group the \xray\ point sources according to the
characteristics of the regions in which they are found.  Finally, we construct
and fit a parameterized XLF model that quantifies how the XLF shape and
normalization evolves as a function of the XRB population age.
Our best-fit model indicates the XRB XLF per unit stellar mass declines in
normalization, by $\sim$3--3.5 dex, and steepens in slope from $\approx$10~Myr
to $\approx$10~Gyr.  
We find that our technique recovers results from past studies of how XRB XLFs
and XRB luminosity scaling relations vary with age and provides a
self-consistent picture for how the XRB XLF evolves with age.

%
\end{abstract}
%


%
\section{Introduction}
%

For decades, it has been known that the \xray\ luminosity from \xray\ binaries
(XRBs) in normal galaxies correlates with optical and far-IR luminosity (e.g.,
Fabbiano \etal\ 1982).  This fact has led to several investigations evaluating
the utility of \xray\ emission as a probe of galaxy properties like
star-formation rate (SFR) and stellar mass ($M_\star$) (see, e.g.,
Fabbiano~2006 for a review; Lehmer \etal\ 2010; Pereira-Santaella \etal\ 2011;
Mineo \etal\ 2012a,b; Boroson \etal\ 2011; Zhang \etal\ 2012).   The emission from the young
($\simlt$100~Myr) high-mass XRB (HMXB; donor-star masses $\simgt$2.5~\msol) population is thought to provide a
relatively unobscured measure of SFR, a quantity that is difficult to determine
well using measurements of the heavily-obscured UV light that is generated by
the most massive, short-lived stars (e.g., Kennicutt \& Evans~2012).  In a complementary way, the
emission from low-mass XRBs (LMXBs; donor-star masses $\simlt$2.5~\msol) may provide an independent tracer of the
star-formation history (SFH), or $M_\star$, due to LMXBs being associated with the
older ($\simgt$1~Gyr) stellar populations.

%
%
\begin{figure*}
\figurenum{1}
\centerline{
\includegraphics[width=18cm]{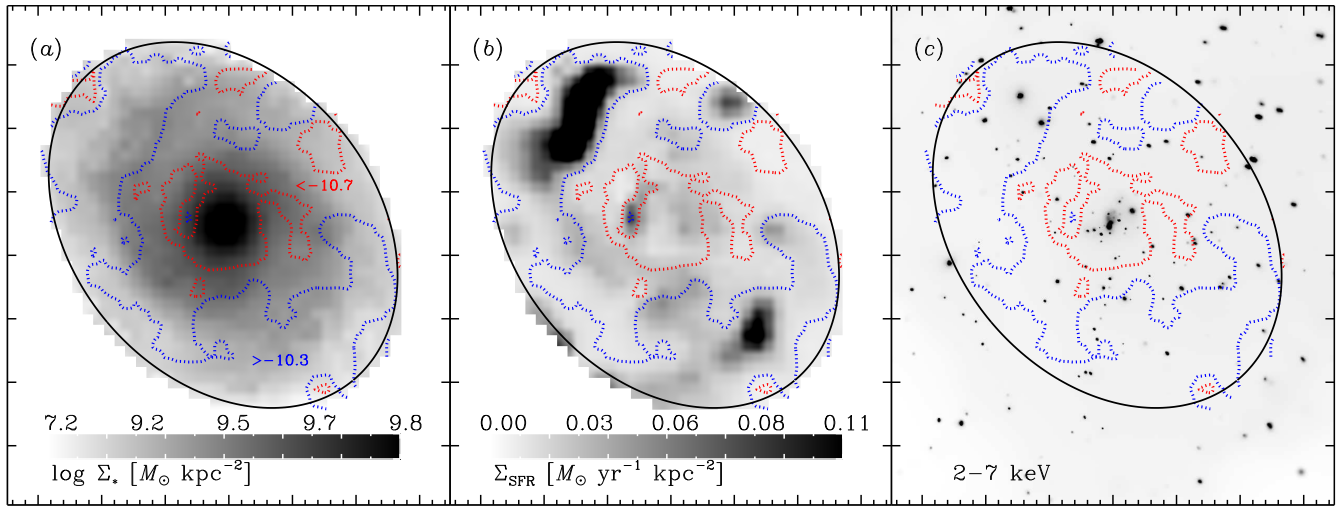}
}
\vspace{0.1in}
\caption{
Stellar mass ($M_\star$), star-formation rate (SFR), and adaptively-smoothed
2--7~keV \chandra\ images of NGC~5194 (M51).  The black ellipse denotes the
$K$-band galaxy size and orientation as derived by Jarrett \etal\ (2003), which
we adopt throughout this paper.  In each panel, we show contours that encircle
regions with SFR/$M_\star > 10^{-10.3}$~yr$^{-1}$ ({\it blue dotted contours}
with ``$>$$-10.3$'' annotation) and SFR/$M_\star < 10^{-10.7}$~yr$^{-1}$ ({\it red
dotted contours} with ``$<$$-10.7$'' annotation).  Regions between these two sets
of contours have SFR/$M_\star = 10^{-10.7}$ to $10^{-10.3}$~yr$^{-1}$.
As expected high-sSFR regions enclose spiral arms, while low-sSFR
regions are primarily located between spiral arms and in the galaxy bulge. 
}
\end{figure*}

There are several obstacles to making XRB emission a well-utilized tracer of
galaxy physical properties.
For one, galaxies that are spatially unresolved in the \xray\ band may suffer
from considerable contamination from low-luminosity active galactic nuclei
(AGN), which can have spectral properties consistent with XRBs.  Furthermore,
in nearby galaxies, where spatial resolution is sufficient to resolve the XRB
population directly (e.g., \chandra\ observations of galaxies at
$\simlt$10--50~Mpc), it is currently not possible to separate confidently the
HMXB and LMXB populations on a source-by-source basis, due to their similar
\xray\ spectral shapes and luminosities.  Finally, the relatively shallow slope
of the SFR-normalized HMXB \xray\ luminosity function (XLF) leads to
frustratingly large statistical variance in the integrated galaxy-wide \xray\
luminosity ($\approx$0.2--0.3~dex; e.g., Gilfanov \etal\ 2004) making SF
estimates highly uncertain.  

Due to the above factors, reliable tracers of physical properties based on, e.g.,
far-UV--to--far-IR spectral energy distributions (SEDs), radio emission,
nebular emission lines (e.g., H$\alpha$ and [O~II]), and/or stellar absorption
features (see, e.g., Madau \& Dickinson~2014 for a recent review) are more
commonly used than \xray\ emission.  However, it has recently become clear that
there are several good reasons to continue to pursue studying and calibrating
the relationships between XRB emission and physical properties, as XRBs can
offer key discriminating information when utilized {\itshape jointly} with data
from other wavelengths.  For instance, XRBs provide a unique probe of the
compact object population, which forms from remnants of $\simgt 8$~\msol\
stars.  As such, the XRB populations are uniquely sensitive to variations in
the high-mass end of the initial mass function (IMF; see, e.g., Peacock \etal\
2014, 2017; Coulter \etal\ 2017).  Also, recent XRB population synthesis
studies predict, and observational constraints appear to confirm, that the
$L_{\rm X}$(HMXB)/SFR and $L_{\rm X}$(LMXB)/$M_\star$ scaling relations are
significantly affected by variations in metallicity and parent stellar
population age, respectively (e.g., Fragos \etal\ 2008, 2013a,b; Kim \&
Fabbiano~2010; Kaaret \etal\ 2011; Basu-Zych \etal\ 2013a, 2016; Brorby \etal\
2014, 2016; Lehmer \etal\ 2014, 2016; Aird \etal\ 2017).  These two quantities (metallicity and age)
are similarly difficult to determine from stellar population synthesis and ISM
fitting of optical/near-IR spectroscopic data, which are currently the most
commonly used methods, and therefore could benefit from complementary and
independent constraints from \xray\ observations.

To advance the use of \xray\ data as a means for constraining galaxy properties
requires empirically calibrating how XRBs evolve with age for a variety of
stellar birth properties (e.g., metallicity) and environmental conditions
(e.g., local stellar densities).  In this paper, we take a preliminary step in
a long-term effort to develop such an empirical calibration by studying how XRB
populations, as characterized by their XLF, vary with stellar age in the nearly
face-on spiral galaxy M51.    Our focus is to study XLFs of XRB populations
that are not affected by extinction and confused by unrelated \xray-emitting
sources (e.g., hot gas and SN remnants).  As such, we restrict our analyses to
\xray\ sources detected in the \hbox{2--7~keV} bandpass, which encompasses only a
fraction of the total number of \xray\ sources detected in the full \chandra\
data set.  A more thorough and complete characterization of all the sources
detected in the rich M51 data set has been summarized in Kuntz \etal\ (2016),
which we refer to throughout this paper.

In our procedure, we first utilize SED fitting of UV--to--far-IR photometric
data to derive SFHs for small subregions of the galaxy.   The details of
this procedure were presented in part~I of this series (Eufrasio \etal\ 2017), and here we make use of the SFH maps that resulted from this work
(see salient features in $\S$2.1)  We then construct a stellar-age dependent
model of the XLF that describes the observed distribution of \xray\ point
sources, including estimates of completeness, contributions from potential
background AGN, and the XRB populations we are modeling.

Throughout this paper, we assume a distance of 8.58~Mpc to M51 based on
measurements of the tip of the red giant branch (McQuinn \etal\ 2016), which
corresponds to an angular scale of 41.6~pc~arcsec$^{-1}$.  All \xray\ fluxes
are corrected for Galactic absorption, assuming a Galactic column density of
$N_{\rm H} = 1.5 \times 10^{20}$~cm$^{-2}$ (Stark \etal\ 1992).  Furthermore,
throughout this paper we make estimates of quantities dependent on the initial
mass functions (IMF), for which we employ the Kroupa~(2001) IMF.

%
\section{Data Analysis}
%

%
%
\begin{figure*}
\figurenum{2}
\centerline{
\includegraphics[width=19cm]{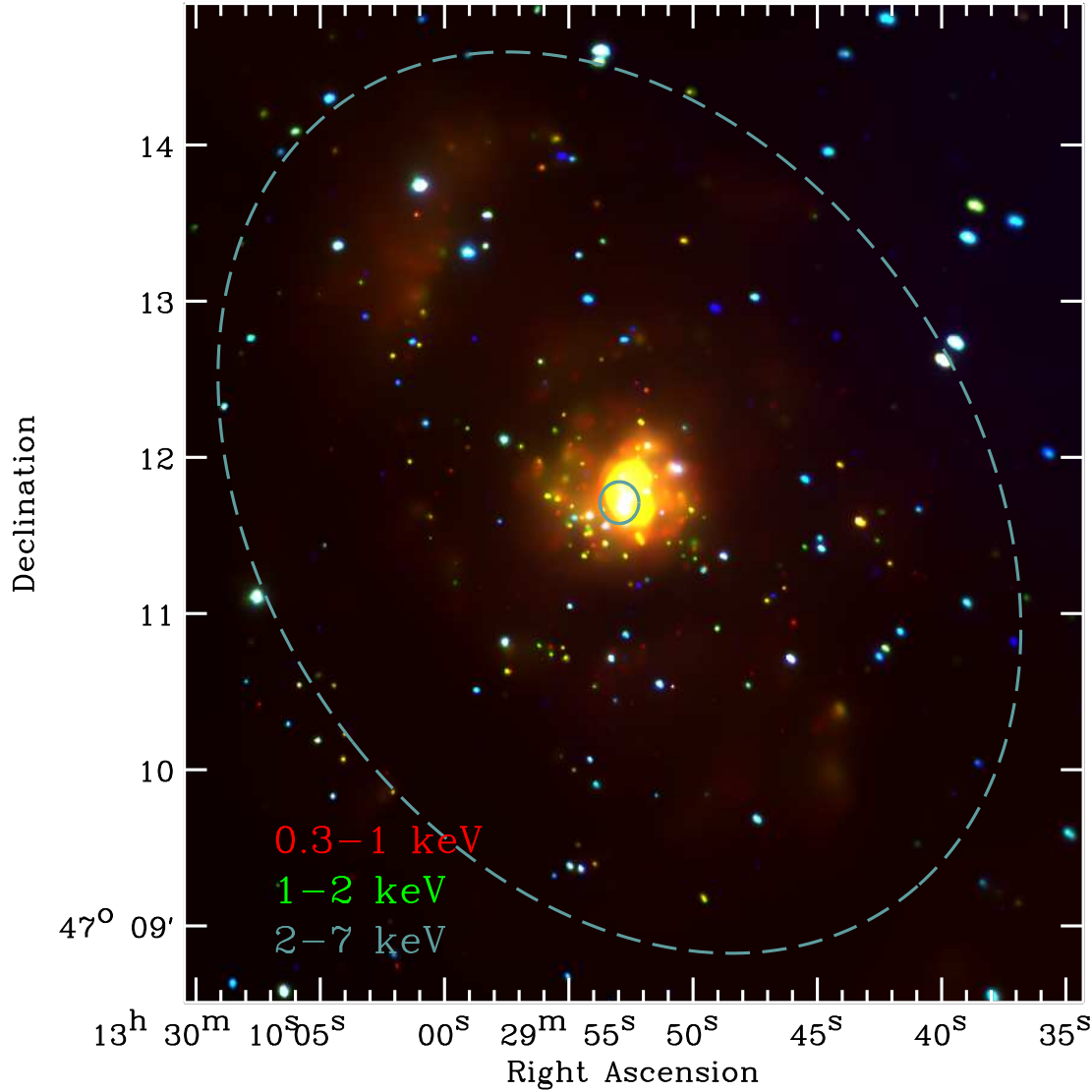}
}
\caption{
Three-color \chandra\ image of M51.  The image was constructed from 0.3--1~keV
($red$), 1--2~keV ($green$), and 2--7~keV ($blue$) exposure-corrected
adaptively smoothed images.  The dashed ellipse represents the total $K$-band
galaxy size and orientation as described in $\S$2.1.  Clear diffuse emission is
evident in the softest band, following the central bulge and spiral arms.  In
our study of the XRB population, we limited our analyses to sources detected in
the 2--7~keV band to avoid confusion with hot gas clumps and other unrelated
soft point sources (e.g., SNe and remnants), and we excluded from consideration
sources within a central circular region of radius 8~arcsec ({\it blue
circle\/}) to avoid the AGN and confused sources.
}
\end{figure*}

\begin{table*}
\begin{center}
\caption{\chandra\ Advanced CCD Imaging Spectrometer (ACIS) Observation Log for M51 (NGC 5194)}
\begin{tabular}{lccccccc}
\hline\hline
 & \multicolumn{2}{c}{\sc Aim Point} & {\sc Obs. Start} & {\sc Exposure Time}$^a$ & {\sc Flaring}$^b$ & {\sc Flaring Time}$^b$ & \\
 \multicolumn{1}{c}{\sc Obs. ID} & $\alpha_{\rm J2000}$ & $\delta_{\rm J2000}$ & (UT) & (ks) & {\sc Intervals} & (ks) & {\sc Obs. Mode}$^c$ \\
\hline\hline
    354\ldots\ldots\ldots\ldots\ldots  &  13      29      49.9 & +47      11       28 &   2000 Jun 20, 08:04     &      \phn\phn14.8   &    \ldots &        \ldots    &   F   \\
    1622\dotfill                       &  13      29      50.1 & +47      11       27 &   2001 Jun 23, 18:47     &      \phn\phn26.8   &    \ldots &        \ldots &  VF   \\
    3932\dotfill                       &  13      29      57.0 & +47      10       39 &   2003 Aug \phn7, 14:32  &      \phn\phn47.5   &    1 &        0.5 &  VF   \\
    12562\dotfill                      &  13      30      04.9 & +47      09       54  &   2011 Jun 12, 06:52     &   \phn\phn\phn9.6   &    \ldots &        \ldots &  VF   \\
    12668\dotfill                      &  13      30      05.4 & +47      09       53 &   2011 Jul \phn3, 10:32  &      \phn\phn10.0   &    \ldots &        \ldots &  VF   \\
    13812\dotfill                      &  13      30      02.2 & +47      10       49 &   2012 Sep 12, 18:25     &         \phn157.5   &    \ldots &        \ldots &   F   \\
    13813\dotfill                      &  13      30      02.2 & +47      10       49 &   2012 Sep 09, 17:48     &         \phn178.7   &    1 &       0.5 &   F   \\
    13814$^d$\dotfill                  &  13      30      03.8 & +47      10       56 &   2012 Sep 20, 07:23     &         \phn189.8   &    \ldots &        \ldots &   F   \\
    13815\dotfill                      &  13      30      04.4 & +47      11       00 &   2012 Sep 23, 08:13     &      \phn\phn67.2   &    \ldots &        \ldots &   F   \\
    13816\dotfill                      &  13      30      04.4 & +47      11       00 &   2012 Sep 26, 05:13     &      \phn\phn73.1   &    \ldots &        \ldots &   F   \\
    15496\dotfill                      &  13      30      03.8 & +47      10       56  &   2012 Sep 19, 09:21     &      \phn\phn41.0   &    \ldots &        \ldots &   F   \\
    15553\dotfill                      &  13      30      05.9 & +47      11       15 &   2012 Oct 10, 00:44     &      \phn\phn37.6   &    \ldots &        \ldots &   F   \\
\rule{0pt}{3ex} 
   Merged$^e$\dotfill                      &  13 30 02.3 & +47 10 54 &      \ldots                &       853.6 &    2 &     1.0   & \ldots    \\
\hline
\end{tabular}
\end{center}
Note.---Links to the data sets in this table have been provided in the electronic edition. \\
$^a$ All observations were continuous. These times have been corrected for removed data that was affected by high background; see $\S$~2.2.\\
$^b$ Number of flaring intervals and their combined duration.  These intervals were rejected from further analyses. \\
$^c$ The observing mode (F=Faint mode; VF=Very Faint mode).\\
$^d$ Indicates Obs.~ID by which all other observations are reprojected to for alignment purposes.  This Obs.~ID was chosen for reprojection as it had the longest initial exposure time, before flaring intervals were removed.\\
$^e$ Aim point represents exposure-time weighted value.
\end{table*}

\subsection{Spectral Energy Distribution Fitting and Property Map Creation}

We determined subgalactic physical properties (e.g., 
SFR, $M_\star$, and SFH) within M51 following the procedure detailed in Eufrasio
\etal\ (2017), part~I of this series.  For full details on this procedure and the data sets used,
we refer the reader to that paper.  Here we describe the salient features of
our analyses and the resulting products that were used in this paper.

First, we utilized a large collection of far-UV--to--far-IR imaging data
(including 16 total bands from \galex, SDSS, 2MASS, \spitzer, and \herschel),
convolved to a common spatial scale, to construct broad band SED maps.
We chose to convolve all images with their respective point
spread functions (PSFs) to a 25~arcsec FWHM spatial resolution.  This is
somewhat more coarse than the FWHM value of the \herschel\ SPIRE 250~$\mu$m
channel, which is the lowest resolution imaging band used in our SED analyses.
The data and resulting product maps were projected
onto a grid with a pixel scale of 10~arcsec ($\approx$400~pc at our adopted
distance).

Next, for each 10\arcsec~$\times$~10\arcsec\ pixel, we performed stellar
population synthesis model fitting using {\ttfamily P\'EGASE} (Fioc \&
Rocca-Volmerange~1997) and a SFH model that consisted
of five discrete time steps, of constant SFR, at \hbox{0--10~Myr}, 10--100~Myr,
0.1--1~Gyr, 1--5~Gyr, and 5--13.6~Gyr.  The stellar emission from these
specific age bins provide comparable bolometric contributions to the SED of a
typical late-type galaxy SFH, and contain discriminating features that can be
discerned in broad-band SED fitting (see Eufrasio \etal\ 2017 for details).  In
our procedure, we fit the SEDs for eight free parameters, including the normalizations
(i.e., SFRs) of all five time steps, as well as three extinction values.
Two of the extinction parameters describe the more heavily extincted population that is present in the
``birth cloud'' immediately following a star-formation event over the
0--10~Myr time frame and a single parameter is used to fit a more characteristic ``diffuse''
component applied to all populations.
The resulting fits thus provide estimates for the SFH in each subgalactic
pixel, which we used to construct the physical parameter maps.
Here we chose to define the value of the SFR as the mean SFR over the last 100~Myr,
which we derive using our SFH results.  This allows us to make equivalent
comparisons with SFR values provided in the literature (e.g., those derived
from scaling relations like those in Kennicutt~1998 and Kennicutt \&
Evans~2012), which are often based on the same assumption.

In Figures~1$a$ and 1$b$, we show the stellar mass and SFR maps derived for M51
with contours of specific SFR (sSFR~$\equiv$~SFR/$M_\star$) indicated.
Throughout this paper, we restrict our analyses to the elliptical region
defined by Mentuch-Cooper \etal\ (2010) for NGC~5194, which is estimated as an
ellipse that traces a nearly constant stellar-mass density of 50~\msol~pc$^{-2}$.  The
ellipse has a semi-major axis of 191.5~arcsec, axis ratio of 0.75, and position
angle of 50~deg east from north; we display the region as a black ellipse in each panel of
Figure~1.  As expected, the most intense star formation, as traced by the sSFR
(enclosed by blue contours in Fig.~1), and the youngest stellar populations are
found in the spiral arms, with less intense star formation and older stellar
populations being located in the galactic bulge and in between the spiral arms.
Going forward, we make use of these maps, along with the spatial locations of
\xray\ point source populations, to statistically constrain how XRB populations
change with environment and evolve over time.

\subsection{Chandra Data Reduction}

All \chandra\ observations (hereafter, ObsIDs) were conducted using the S-array
of the Advanced CCD Imaging Spectrometer (ACIS-S; see Table~1 for full
observation log).  Given the major-axis length of the M51 disk
($\approx$6.4~arcmin), the full galactic extent is covered by the single
ACIS-S3 chip in nearly all 12 ObsIDs.  For our data reduction, we made use of
{\ttfamily CIAO}~v.~4.8 with {\ttfamily CALDB}~v.~4.7.1.\footnote{http://cxc.harvard.edu/ciao/}  We began by
reprocessing the pipeline produced events lists, bringing level~1 to level~2
using the script {\ttfamily chandra\_repro}.  The {\ttfamily chandra\_repro}
script runs a variety of {\ttfamily CIAO} tools that identify and remove events
from bad pixels and columns, and filter the events list to include only good
time intervals without significant flares and non-cosmic ray events
corresponding to the standard \asca\ grade set (\asca\ grades 0, 2, 3, 4, 6).

Using the reprocessed level~2 events lists for each ObsID, we generated
preliminary 0.5--7~keV images and point-spread function (PSF) maps (using the tool
{\ttfamily mkpsfmap}) with a monochromatic energy of 1.497~keV and an encircled
counts fraction (ECF) set to 0.393.   For each ObsID, we created preliminary
source catalogs by searching \hbox{0.5--7~keV} images with {\ttfamily
wavdetect} (run including our PSF map), which was set at a false-positive
probability threshold of $1 \times 10^{-5}$ and run over seven wavelet scales
from 1--8 pixels (1, $\sqrt{2}$, 2, 2$\sqrt{2}$, 4, 4$\sqrt{2}$, and 8).  To
measure sensitively whether any significant flares remained in our
observations, we constructed point-source-excluded \hbox{0.5--8~keV} background
light curves for each ObsID with 500~s time bins.  We found two 500~s intervals
across all ObsIDs with flaring events of $\simgt$3~$\sigma$ above the nominal
background; these intervals were removed from further analyses and the
resulting flare-free exposures are presented in Table~1.

For each ObsID, we used the preliminary source catalogs to register each
flare-free aspect solution and events list to ObsID:13814, which had the
longest exposure time.  This process was carried out using {\ttfamily CIAO}
tools {\ttfamily reproject\_aspect} and {\ttfamily reproject\_events},
respectively.  The astrometric reprojections resulted in very small linear
translations ($<$0.38~pixels; $<$0\farcs19), rotations ($<$0.1~deg), and stretches ($<$0.06\%
of the pixel size) for all ObsIDs.  We created a merged events list, as well as
a series of images in a variety of bands (including 0.3--1~keV, 1--2~keV,
0.5--2~keV, 2--7~keV, and 0.5--7~keV) and monochromatic exposure maps (at
1.497, 4.51, and 2.53~keV; in units of s~cm$^2$), using the {\ttfamily CIAO}
script {\ttfamily merge\_obs}.  The {\ttfamily merge\_obs} script properly
projects all ObsID events lists and exposure maps to a common reference frame,
and generates projected products for each ObsID that correspond to the merged
products.  We converted the exposure maps of each ObsID into vignetting-corrected
exposure-time maps (in units of s) by dividing the exposure maps by the maximum
effective area in that ObsID (see Hornschemeier \etal\ 2001 and Zezas \etal\ 2006 for additional
details).  These exposure-time maps were subsequently merged to form a merged
vignetting-corrected exposure-time map.  We further created 90\% enclosed
count-fraction PSF maps for each of the ObsIDs for an energy of 4.51~keV.  We
then created a vignetting-corrected exposure-time weighted merged PSF map, by
summing the product of the PSF and exposure-time maps for each ObsID and dividing by
the merged exposure-time map.

\subsection{Main Catalog Creation and Properties}

In Figure~2, we show a three-color (0.3--1~keV, 1--2~keV, and 2--7~keV)
adaptively-smoothed image of M51, with the adopted elliptical boundary
highlighted (see $\S$2.1 for details).  Within the adopted galaxy boundaries, there are 208 significantly detected \xray\ point sources (from the Kuntz \etal\ 2016 catalogs) and clear diffuse emission in the bulge region
and along the spiral arms.  A variety of \xray\ point source types are present,
including supernovae and remnants, XRBs, background AGN, and clumps of hot gas.
Given that the goal of our study is to characterize the XRB populations, we
limit ourselves to point sources detected in the 2--7~keV band (Fig.~1$c$).  While this
imposed limitation reduces the sensitivity of the \chandra\ survey to XRBs
overall, it has the benefit of providing a cleaner sample of the XRB population
by excluding ``soft'' unrelated sources like supernovae and remnants and hot
gas clumps.

We constructed a point source catalog by searching the \hbox{2--7~keV} image
using {\ttfamily wavdetect} at a false-positive probability threshold
of $1 \times 10^{-6}$ over the $\sqrt{2}$ sequence (see above).  We ran
{\ttfamily wavdetect} using the timing map and 90\% enclosed-count fraction PSF
map, which resulted in a source catalog with properties (e.g., positions and
counts) appropriate for point sources.  We inspected all the images and source
regions by eye to see if any additional source candidates were missed or
confused.  In particular, in the crowded nuclear regions of the galactic center,
low-flux point sources may not be picked up by {\ttfamily wavdetect} due to the
increased backgrounds.  We found that point-source crowding in the central region
was prohibitively large to obtain accurate point-source properties.  As such,
we excluded from further analyses a circular region at the center of the galaxy
with an 8~arcsec radius.  In total, 86 point sources were detected in the
\hbox{2--7~keV} band that were within the galactic footprint, yet outside the
8~arcsec radius central exclusion area.  Hereafter, we exclusively utilize these sources and
the adopted {\ttfamily wavdetect} catalog, which we define as our {\it main catalog}.

Given that we made use of the vignetting-corrected timing maps described above
when running {\ttfamily wavdetect}, our main catalog contained measurements of
the vignetting-corrected source count rates, which are background subtracted.
In order to convert source \hbox{2--7~keV} count rates to \hbox{2--10~keV}
fluxes, we used the {\ttfamily CIAO} script {\ttfamily specextract} to
construct an exposure-weighted response matrix file (RMF) and ancillary
response file (ARF) appropriate for a hypothetical source located at the
exposure-weighted aim point.  In {\ttfamily xspec}
v.~12.8.2\footnote{https://heasarc.gsfc.nasa.gov/xanadu/xspec/}, we used the
{\ttfamily fakeit} command to create a fake source with a power-law model that
has Galactic absorption (see $\S$1) and photon-index $\Gamma = 1.7$ and derived
the count-rate to flux conversion factor for such a source.  We utilized this
same count-rate to flux conversion factor for all sources within our main
catalog, since the detailed spectral shape of faint sources is uncertain for the majority of the sources in our catalog.
Our adopted X-ray spectral model is broadly appropriate for the 2--10~keV
emission from X-ray binaries across a variety of X-ray luminosities and
describes well the overall spectral shapes of a variety of XRB-dominated nearby
galaxies (e.g., Mineo \etal\ 2012; Lehmer \etal\ 2014, 2015).  For the sources with
$>$300 2--7~keV counts, we performed basic spectral fitting and found a mean
photon index of $\Gamma = 1.69$, albeit with a large scatter of $\sigma_\Gamma
= 0.54$.  This translates into an estimated error on the luminosity of
$\approx$17\% due to the assumed SED, which is only somewhat larger than the
$\approx$12\% median error on photon statistics for all sources within M51.

The faintest sources in our main catalog have $\approx$6--10 net counts in the
2--7~keV band, which corresponds to a 2--10~keV flux limit of
$\approx$(9--19)~$\times 10^{-16}$~\flux.  At the distance of M51, these
sources would have 2--10~keV luminosities of $\approx$(1.1--2.1)~$\times
10^{36}$~\lum.  Hereafter, unless stated otherwise, we refer to \xray\
luminosities $L_{\rm X}$ as pertaining to the \hbox{2--10~keV} band.

We compared our main catalog sources with those provided in Kuntz \etal\
(2016).  Out of the 86 sources in our catalog, we found matches to all but
three sources: J132952.2+471100, J132954.3+471153, and J132943.4+471201.  These
three sources are faint, but clearly detected with \hbox{12--21} net counts in
the 2--7~keV band.  Kuntz \etal\ (2016) thoroughly searched multiple bands for
source candidates, and eliminated several candidates based on their
significance against the background and source extension, as derived by
{\ttfamily ACIS EXTRACT} (Broos \etal\ 2012).  Visual inspection of the
lower-energy band images revealed that the three sources unique to our main
catalog were located in regions with strong $\simlt$1~keV emission from hot
gas, and were thus likely to be rejected based on the extended emission from
the nearby hot gas and the use of {\ttfamily ACIS EXTRACT} for evaluating
source significance.  For the remaining 83 sources, we found that 12 had
classifications by Kuntz \etal\ (2016) based on broadband and H$\alpha$ \hst\
imaging.  Of these twelve sources, four were background galaxies, four were
directly associated with star-forming regions, and six were coincident with
H$\alpha$ bubbles.  The latter six sources were noted to be a combination of
transient and compact sources.  Taken together, these classifications suggest
that the majority of our sources are consistent with being XRBs.  Our modeling
techniques, discussed below, account statistically for potential background
sources, which we estimate to be $\approx$6--10 over the footprint of M51,
which suggests a reasonable fraction of the background sources may be
identified by Kuntz \etal\ (2016).  However, given the complex selection
function of background sources through a face-on spiral galaxy like M51, we do
not attempt to remove the few known background sources in our procedure below.

%
\section{Analysis and Results}
%

The primary goals of this paper are to (1) decompose the relative contributions
of HMXBs and LMXBs to the total XRB luminosity by using subgalactic properties
and (2) construct a comprehensive model for how the XRB XLF evolves as a function of
age that agrees with empirical constraints on the observed XLF and subgalactic
SFHs.  In the sections below, we describe in detail our
procedure for accomplishing these goals.

\subsection{Completeness Estimates}

To compute XLFs, we made use of the \hbox{2--7~keV} catalog of 86
sources in the main catalog presented in $\S$2.  Spatial variations in \xray\
sensitivity, due to local background fluctuations, PSFs, effective exposures
(e.g., chip gaps and bad pixels and columns), and source crowding, combined
with incompleteness, can have significant effects on the shape of the {\it
observed} XLF at \xray\ luminosities within a factor of $\approx$10 of the
detection limits.  Therefore, these effects need to be properly characterized
to compute the XLF.

In our analyses, we made use of the {\it observed} XLFs and models of the {\it
intrinsic} XRB XLFs convolved with completeness functions that we derived from a Monte
Carlo procedure.  Our procedure was designed to provide the recovery fraction
as a function of net source counts and source position.  We began by
constructing a series of 3,000 mock images.  Each image consisted of our
original 2--7~keV image plus 49 fake sources added to the image.  Each source
location was chosen to lie within the boundaries of a single box, with the
fake images containing a grid of $7 \times 7$ total boxes (defined in equal
intervals of right ascension and declination).  For a given image, only one
source was placed in each box with a random location within that box.  For each simulation, we pre-defined
the total number of counts that would actually be registered on the detector,
and created 200 mock images for 15 different choices of net counts (spanning
3--260 counts).  The source counts were added to the fake images one at a time,
with the distribution of photons following the average PSF shape, which we
determined using the count distributions from bright sources within the image.

We note that the above procedure for creating fake images is somewhat different
from standard procedures available through the {\ttfamily marx}\footnote{See
http://space.mit.edu/cxc/marx/.} ray-tracing code.  We chose our approach
because it very quickly allows us to construct many fake images with fake
sources that contain an exact number of net counts that we define.  We
performed several curve-of-growth tests of the fake sources and verified that
the count distributions from our approach are consistent with the PSFs of real
sources in the image.  Given the relatively small extent of M51, the PSF across the galactic footprint
does not have significantly distorted shapes (i.e., non-circular), as is known to
be the case for the \chandra\ PSF at large off-axis angles.  As such, we
caution that this approach will not likely work for sources with very large
off-axis angles and distorted PSFs.

To construct completeness functions, we repeated the source detection procedure
described in $\S$2 for all 3,000 mock images and compared mock catalogs with
the input catalogs.  In Figure~3, we show the fraction of fake sources
recovered as a function of counts and angular distance from the M51 galactic
center.  We find the highest levels of completeness for regions near the center
of the galaxy and only a mild decline in completeness in the outer regions of
the galaxy $\approx$3~arcmin away from the galactic center.  Such variations
are expected due primarily to the larger average PSF size with galactocentric
distance.  In $\S$3.2 below, we describe how we use our completeness functions
when measuring XRB XLFs.

%
%
\begin{figure}
\figurenum{3}
\centerline{
\includegraphics[width=9.2cm]{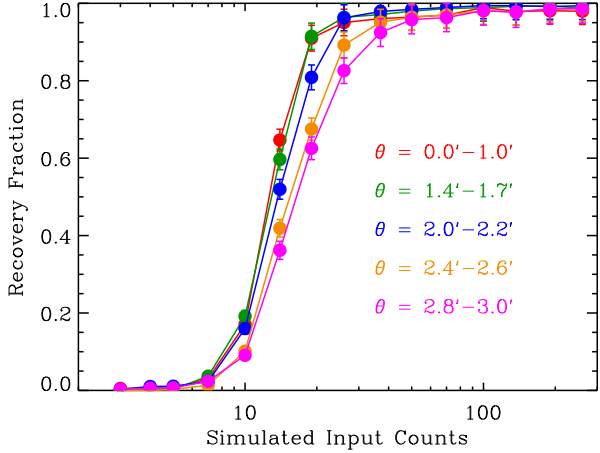}
}
\caption{
Fraction of simulated sources recovered as a function of counts for five M51
galactocentric distances (see annotations).  These recovery fraction curves
show the highest completeness for sources near the center of the galaxy with
declining completeness at larger galactocentric distances, where the average
PSF is largest.
}
\end{figure}

%
%
\begin{figure*}
\figurenum{4}
\centerline{
\includegraphics[width=9cm]{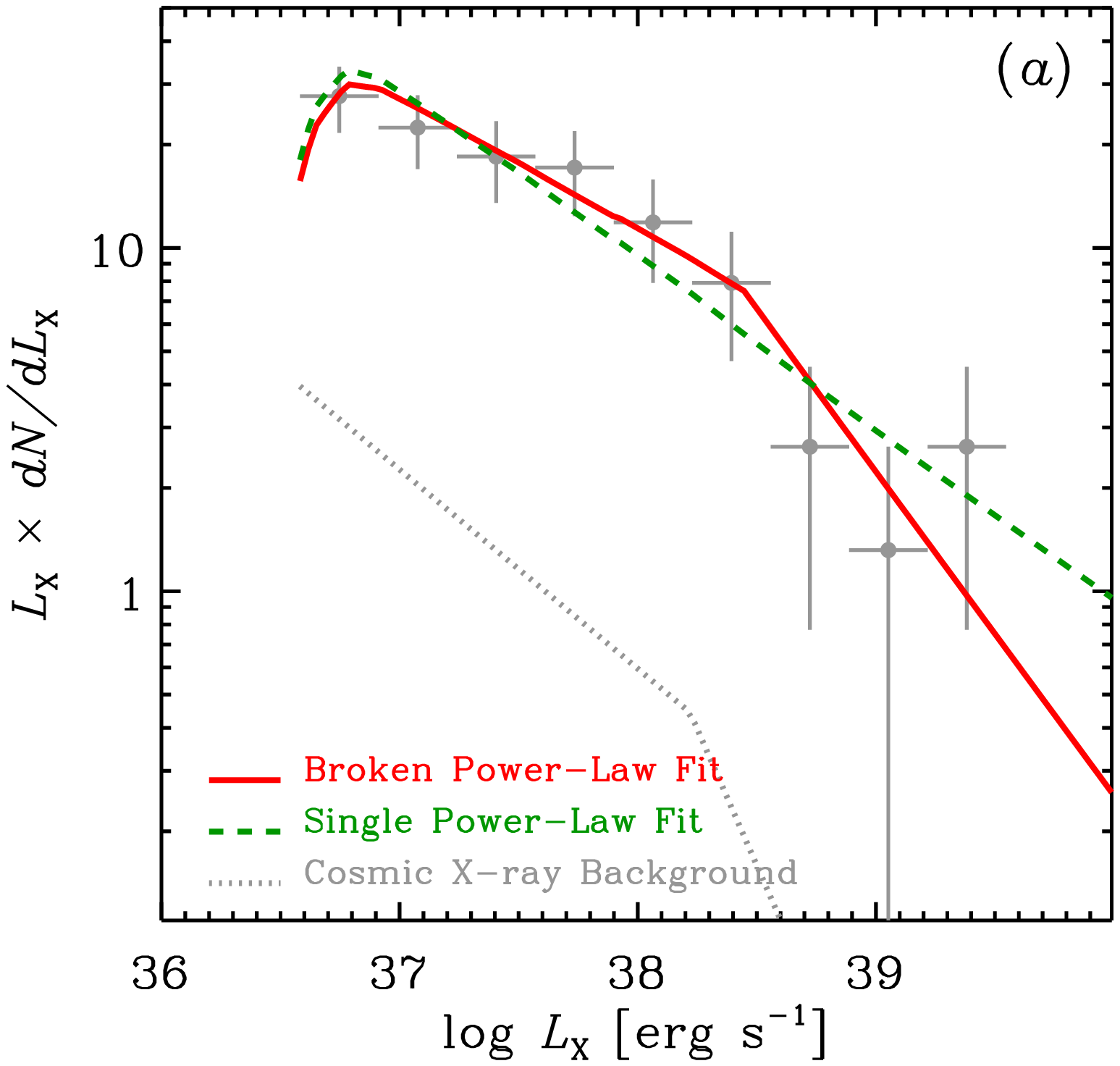}
\hfill
\includegraphics[width=9cm]{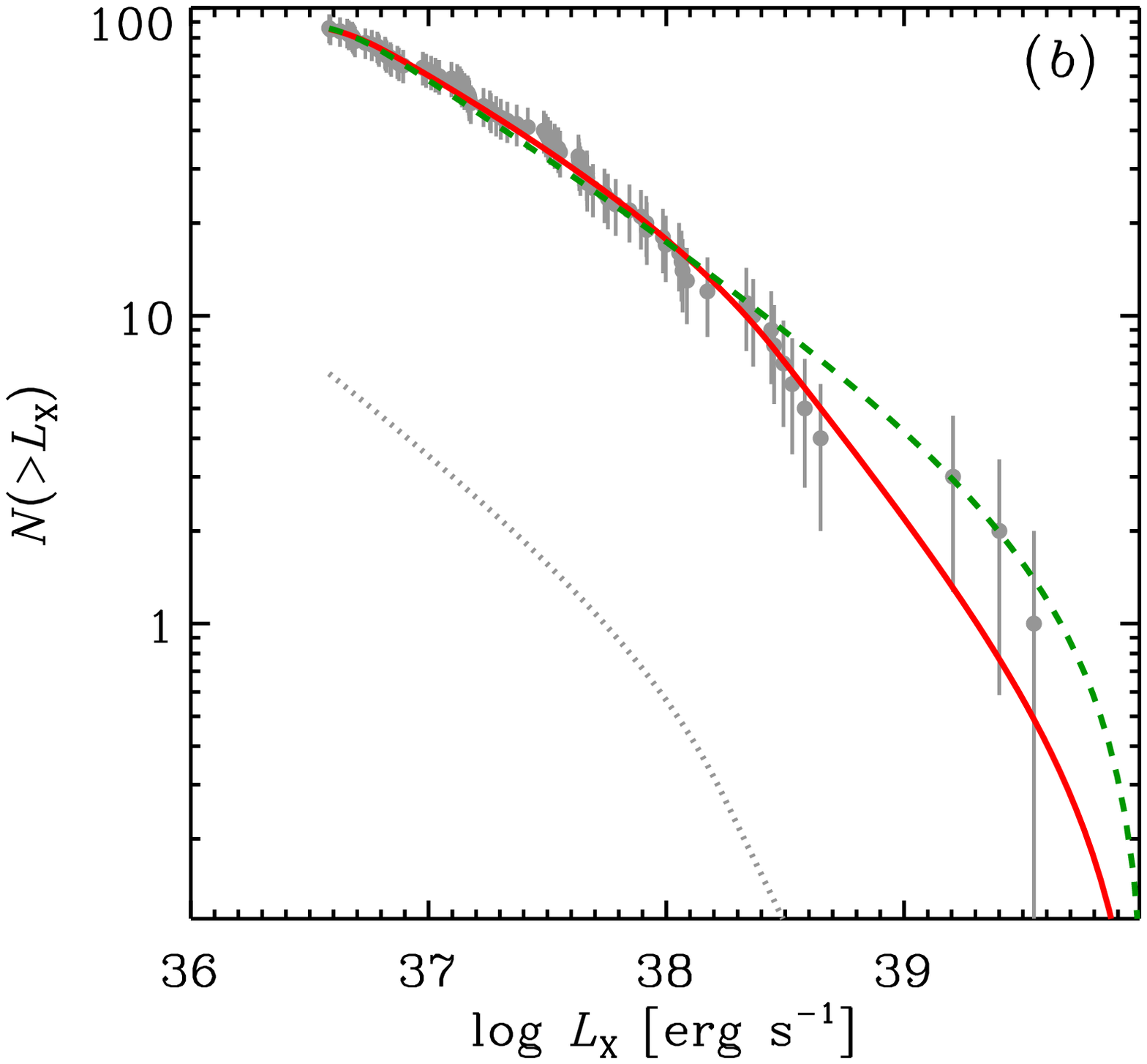}
}
\caption{
Galaxy-wide differential ($a$) and cumulative ($b$) XLFs for \xray\ point
sources in M51 ({\it gray filled circles with 1$\sigma$ Poisson errors\/}).  Note that the
differential XLF plotted here is for display purposes, and does not show the
finer binning used in fitting to the data.  Best-fit models that include
expected CXB sources plus XRBs, folded in with estimates of incompleteness (as
described in $\S$3.2), have been shown for a broken power-law fit ({\it solid
red curves\/}) and a single power-law fit ({\it dashed green curves\/}).  The
data are shown as gray circles with error bars, and have {\it not} been
corrected for incompleteness and background sources.  Incompleteness only
affects the faintest source population below 2--10~keV luminosities of $L_{\rm
X} \approx 10^{37}$~\lum.  Both the broken and single power-law models
provide good fits to the data based on {\ttfamily cstat}.
}
\end{figure*}

%
\subsection{Galaxy-Wide XRB XLF of M51}

We began our XLF analyses by computing the galaxy-wide 2--10~keV XLF for M51.
In Figure~4, we display the galaxy-wide {\it observed XLF} for M51 in both
differential and cumulative form ({\it gray filled circles with 1$\sigma$
Poisson error bars\/}).  These data contain completeness-uncorrected
contributions from both XRBs in the disk of M51, as well as background \xray\
point sources from the cosmic \xray\ background (CXB; e.g., Kim \etal\ 2007;
Georgakakis \etal\ 2008).
In principle, there may also be \xray\ point sources associated with foreground
Galactic stars; however, inspection of optical counterparts to the \xray\
sources, and characterizations provided by Kuntz \etal\ (2016), do not yield
any Galactic stellar candidates.  

We fit the observed galaxy-wide XLF following a forward-fitting approach, in
which we include contributions from the XRBs and CXB sources, with
incompleteness folded into our models.  For the {\it intrinsic} XRB XLF, we
attempted both single and broken power-law models of the respective forms:
$$\frac{dN}{dL_{\rm X, 38}} = K_s \;f(L_{\rm X}),$$
\begin{equation}
\begin{split}
f(L_{\rm X}) \equiv \left \{ \begin{array}{lr}
L_{\rm X, 38}^{-\gamma}  & \;\;\;\;\;\;\;\;(L_{\rm X, 38} < L_c) \\ 0,  &
(L_{\rm X, 38} \ge L_c) \\ \end{array}
  \right.
\end{split}
\end{equation}

$$\frac{dN}{dL_{\rm X, 38}} = K_b \;g(L_{\rm X}),$$
\begin{equation}
\begin{split}
g(L_{\rm X}) \equiv \left \{ \begin{array}{lr}
L_{\rm X, 38}^{-\alpha_1}  & (L_{\rm X, 38} \le L_{\rm b}) \\ L_{\rm
b}^{\alpha_2 - \alpha_1}L_{\rm X,38}^{-\alpha_2}, & \;\;\;\;(L_{\rm b} < L_{\rm
X, 38} < L_c)\\ 0,  & (L_{\rm X, 38} \ge L_c) \\ \end{array}
  \right.
\end{split}
\end{equation}
where $L_{\rm X, 38}$ is the 2--10~keV \xray\ luminosity in units of
$10^{38}$~\lum, $K_s$ and $K_b$ are normalization terms at $L_{\rm X, 38} = 1$
($L_{\rm X} = 10^{38}$~\lum), $\gamma$ is the single power-law slope, and for
the broken power-law, $\alpha_1$ is the faint-end slope, $L_{\rm b}$ is the
break luminosity (in units of $10^{38}$~\lum), and $\alpha_2$ is the bright-end
slope.  Both functions were terminated at a cut-off luminosity of $L_c = 100$,
which we adopted based on the literature (e.g., Mineo \etal\ 2012).  In
Equations~(1) and (2) we also defined the functions $f(L_{\rm X})$ and
$g(L_{\rm X})$ as short-hand for the $L_{\rm X}$-dependent portions of the
single and broken power-laws, respectively.

For the CXB contribution, we utilized a fixed form from the number-counts
estimates provided by Kim \etal\ (2007).  The Kim \etal\ (2007) extragalactic
number counts provide estimates of the number of sources per unit area versus
2--8~keV flux.  The best-fit functions follow a broken power-law distribution
with parameters derived from the combined \chandra\ Multiwavelength Project
(ChaMP) and \chandra\ Deep Field-South (CDF-S) extragalactic survey data sets
(see Table~4 of Kim \etal\ 2007).  The number counts were converted to observed
\hbox{2--10~keV} XLF contributions by (1) multiplying the number counts by the
areal extent of M51 as defined in $\S$2.1 (24.0~arcmin$^2$); (2) converting CXB model fluxes to \xray\
luminosities, given the distance to M51; and (3) multiplying the luminosities by a small
correction factor to bring the 2--8~keV band luminosities to our adopted
2--10~keV band.  The fixed CXB contribution is shown in Figure~4 as dotted curves;
however, we note that these curves are not corrected for completeness (see
below).

A complete model of the observed XLF, $dN/dL_{\rm X}({\rm obs})$, consists of
the XRB intrinsic XLF component, $dN/dL_{\rm X}({\rm XRB})$, from Equation~(1) or (2),
plus the fixed CXB curve, $dN/dL_{\rm X}({\rm CXB})$, convolved with a
galaxy-wide weighted completeness correction, $\xi(L_{\rm X})$, which was
constructed using the radial-dependent completeness
estimates calculated in $\S$3.1.  $\xi(L_{\rm X})$ was
thus calculated by statistically weighting the contributions from the model XLF at
each annulus according to the observed distributions of \xray\ point sources.
Formally, we computed $\xi(L_{\rm X})$ using the following relation:
\begin{equation}
\xi(L_{\rm X}) = \sum_i f_{\rm recov, i}(L_{\rm X}) \times w_i,
\end{equation}
where $f_{\rm recov, i}(L_{\rm X})$ is the recovery-fraction curve for the
$i$th annular bin (see Fig.~3) and $w_i$ is the fraction of total number of
galaxy-wide sources within the $i$th annuluar bin based on the observed
point-source distributions.

We thus modeled the observed XLF using a multiplicative
model
\begin{equation}
dN/dL_{\rm X}({\rm obs}) = \xi(L_{\rm X}) [dN/dL_{\rm X}({\rm XRB}) +
dN/dL_{\rm X}({\rm CXB})].
\end{equation}
In practice, we constructed the observed $dN/dL_{\rm X}({\rm obs})$ using a
small constant bin of $\delta \log L_{\rm X} = 0.036$~dex spanning the minimum luminosity
of the subsample ($L_{\rm X} = 2.8 \times 10^{36}$~\lum) to a maximum $L_{\rm X} = 10^{40}$~\lum.  Therefore the
majority of bins contained zero sources up to a maximum of three sources per
bin.  We evaluated the goodness of fit for our double power-law models using
the Cash statistic ({\ttfamily cstat}; Cash~1979).  Our single and broken
power-law models contained two ($K_s$ and $\gamma$) and four ($K_b$,
$\alpha_1$, $L_b$, and $\alpha_2$) free parameters, respectively (see Eqn.~(1)
and (2)).  For the broken power-law model, we chose to 
constrain the break luminosity to the $L_b =$~2--5 range to avoid confusion between
solutions that place the break luminosity near either end of the luminosity
range of the \xray\ sources.  This choice was based on visual inspection of the
left panel of Figure~4 and is further motivated by observations in the literature of a break near the luminosity (see below).

Best-fit parameters for our models were determined by minimizing {\ttfamily
cstat}.  In this procedure, we found best fit solutions using custom software,
which implemented a Monte Carlo chain of perturbing the variables randomly
10,000 times around successive best-fit solutions until convergence.  We tested the convergence
of this procedure by using very different combinations of initial parameter
guesses, but found robust convergence in all tests.  Once a best-fit solution
was isolated, we constructed multi-dimensional grids of parameter values around
the best-fit solutions and calculated the probability-density spaces in the
vicinity of the best solutions.  In Figure~4, we show the best-fitting observed
galaxy-wide XLF models for the single (green curves) and broken (red curves)
power-law fits in both the differential and cumulative form.  The best-fit
parameters, and their 1$\sigma$ confidence errors, are tabulated in Table~2.  

%
%
\begin{figure*}
\figurenum{5}
\centerline{
\includegraphics[width=18cm]{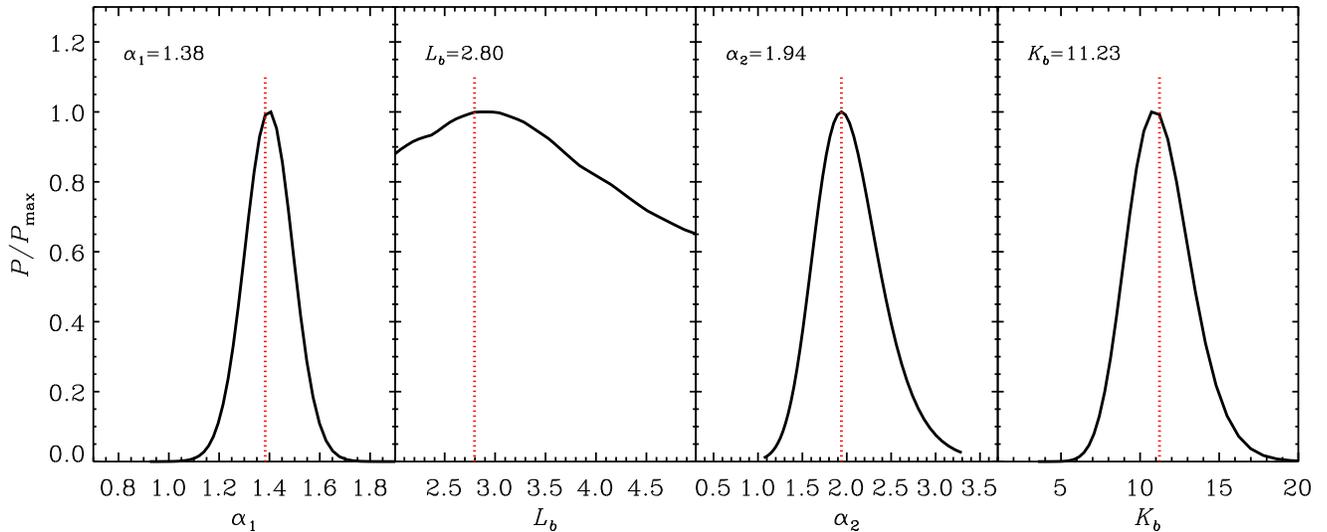}
}
\caption{
Marginalized probability distribution functions for the broken power-law
parameters that were fit to the galaxy-wide XLFs.  The parameters include the
faint-end power-law slope $\alpha_1$ ($a$),  break luminosity $L_b$ ($b$),
bright-end power-law slope $\alpha_2$ ($c$), and normalization to the XLF $K_b$
($d$) (see Eqn.~(2)).  For reference, the best-fit values are
highlighted with vertical red dotted lines.
}
\end{figure*}

We assessed the goodness of fit for our models by performing simulations.  We
constructed 50,000 simulated XLFs that are taken to be statistical ``draws''
from the best-fit XLF.  For example, for the case of a single power-law, one of
our simulated XLFs will be constructed by (1) perturbing the total number of
\xray\ point sources predicted from the model in a Poisson manner, and (2)
assigning $L_{\rm X}$ values to the sources probabilistically following the
best-fit XLF solution given in Table~2.  For this set of simulated data, we
then calculated the {\ttfamily cstat} value assuming the input model.  The
distribution of {\ttfamily cstat} values provides a measure of the probability
of obtaining a given {\ttfamily cstat} value and allows us to assess whether a
given data set is consistent with being drawn from the model.  We find that
both the single and broken power-law models provide good fits to the data, with
the probability of obtaining the measured {\ttfamily cstat} values, or larger,
being $P(\ge${\ttfamily cstat}$) =$~0.75 and 0.66, respectively.   As such,
a power-law break in the XLF is not formally required in the overall \xray\
point-source population in M51.

In Figure~5, we show the marginalized probability density distribution
functions for the broken power-law fit parameters $\alpha_1$, $L_b$,
$\alpha_2$, and $K_b$ with the best-fit value annotated.  All parameters, with
the exception of $L_b$, are well constrained.  $L_b$ itself shows a distinct
maximum likelihood around $L_b \sim 2$ within the range of values explored.
This value is consistent with high-luminosity breaks reported in the past (see,
e.g., Sarazin \etal\ 2000; Gilfanov~2004; Kim \& Fabbiano~2004; Zhang \etal\
2012), and is often explained as being associated with a transition to almost
exclusively black hole accretors, since the Eddington luminosity of a neutron
star is near this limit.  We did find that another peak of comparable, but
somewhat lower probability around $L_b \sim 0.2$ when allowing $L_b$ to vary
outside of this range, and this solution likely represents an additional real
break, which has been reported in past studies as well (e.g., Gilfanov~2004;
Voss \& Gilfanov~2006; Voss \etal\ 2009; Zhang \etal\ 2012).  The nature of this break is
more mysterious, but may be associated with a reduction in XRBs with main
sequence donor stars at lower luminosities and the onset of XRBs with red-giant
donors at higher luminosities (e.g., Fragos \etal\ 2008).  

\subsection{LMXB and HMXB XLF Decomposition}

As with most spiral galaxies, the stellar populations within M51 span a wide
range of stellar ages due to a sustained SFH spanning several Gyr.  In the case
of M51, the most active episodes of star formation occurred more than 100~Myr
ago, with the most active growth occuring over the 0.1--5~Gyr time frame (see,
e.g., Eufrasio \etal\ 2017 and references therein).  Inevitably, the XRB
population within M51, and late-type galaxies in general, will contain both
LMXB and HMXB populations.  Although there have been several investigations of
the stellar-mass scaling of LMXB XLFs using elliptical galaxies (e.g.,
Gilfanov~2004; Zhang \etal\ 2012; Lehmer \etal\ 2014; Peacock \etal\ 2014, 2017) and
the SFR scaling of HMXB XLFs based on spirals (e.g., Grimm \etal\ 2002; Mineo
\etal\ 2012), these studies assume that either LMXBs or HMXBs dominate the
observed XRB populations.  However, there are reasons to believe that such
assumptions are unlikely to be fully correct.  For example, XRB population
synthesis models predict that the LMXB XLF normalization declines significantly
with increasing stellar population age (e.g., Fragos \etal\ 2008), suggesting
that stellar-mass scaled LMXBs based on elliptical galaxies alone are likely to
underpredict the LMXB XLF in young galaxies.  Some evidence for this has
already been apparent in the LMXB populations of ellipticals with varying mean
stellar ages (see Kim \etal\ 2009; Lehmer \etal\ 2014), and there is some
suggestion that the younger mean stellar population within M51 itself may be
influencing the LMXB XLF (see Kuntz \etal\ 2016 and below).

In this section, we make use of the $M_\star$ and SFR maps presented in
Figure~1 (see Eufrasio \etal\ 2017 for details) as a means for
probabilistically separating, respectively, LMXB and HMXB contributions to the
XLF.  Our strategy assumes that the normalizations (not the shapes) of the LMXB
and HMXB XLFs scale with $M_\star$ and SFR, respectively, on scales down to $\sim$1--2~kpc.
Such an assumption may not be fully accurate, in particular, due to the
influence of XRB natal kicks, in which SNe that precede the formation of the
compact object within the XRB can lead to a strong peculiar velocity
(\hbox{$\sim$100--200~km~s$^{-1}$}) of the binary system relative to its birth
population (e.g., Brandt \& Podsiadlowski~1995).  Thus far, empirical studies
indeed show that these natal kicks are likely to have some influence on the
distribution and velocities of LMXBs in the Milky Way relative to their parent
stellar populations (e.g., Podsiadlowski \etal\ 2005; Repetto \etal\ 2012,
2017; Maccarone \etal\ 2014).  However, such kicks are estimated to scatter
only a fractionally small number of systems relative to their parent stellar
population on the subgalactic scales that we probe here (typically $\sim$2
kpc), as evidenced by only a small excess of sources found outside of
elliptical galaxies and the observation that scaling relations appear to hold
on local scales (e.g., Kundu \etal\ 2007; Zhang \etal\ 2013; Mineo \etal\
2014).  For young HMXBs, the typical center-of-mass velocity of
$\simlt$30~km~s$^{-1}$ is not enough to significantly displace these objects
from their parent population over the binary lifetimes (see Antoniou \& Zezas
2016).  For LMXBs, the spatial distribution of the stellar populations from
which they are born are smoothed out by the galaxy velocity field, in the same
way as the LMXBs themselves.  Therefore, the LMXBs are sampling the same
average old stellar populations.  To this extent, we expect that XRBs found in
areas with the highest sSFR will have the largest likelihood of being HMXBs,
while those in the lowest sSFR will have the highest likelihood of being LMXBs.
The probability of the \xray\ source being a background \xray\ source from the
CXB is estimated following the number counts and the areas enclosed by a given
sSFR range (see $\S$3.2 for details).

%
%
\begin{figure*}
\figurenum{6}
\centerline{
\includegraphics[width=18cm]{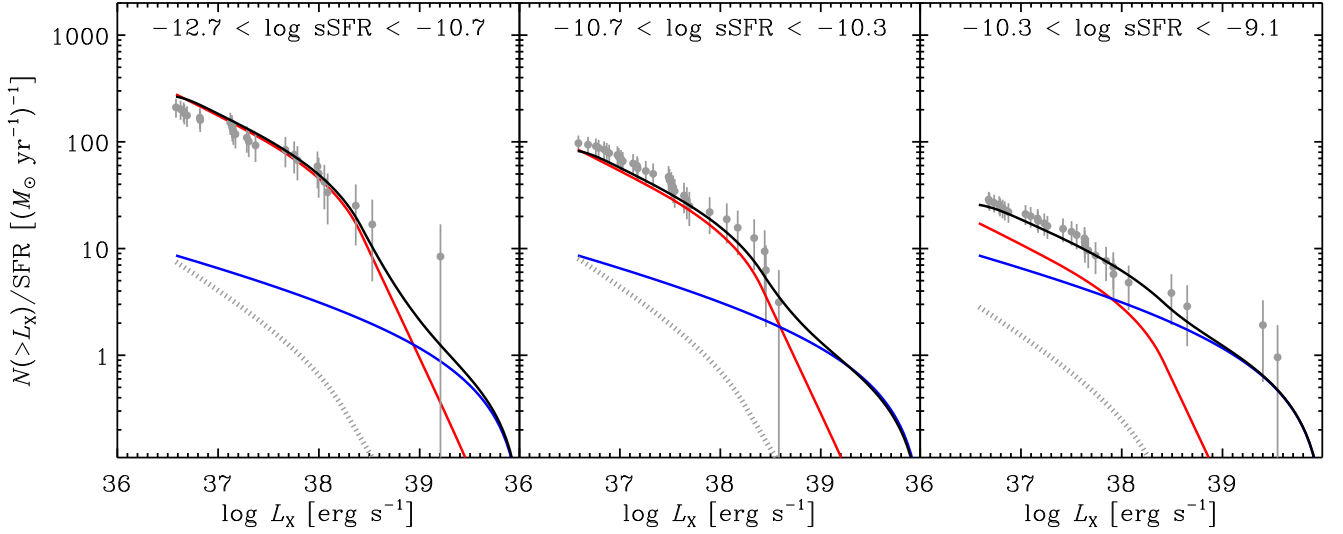}
}
\vspace{0.1in}
\caption{
SFR-normalized XLFs for three subregions of M51 that have been separated by
sSFR (see annotations at the tops of each panel).  As per Fig.~4, gray filled circles with error bars are the measured XLFs without corrections and the dotted curves represent the estimated CXB contributions. There is clear evidence that
the observed XLF both declines in normalization and flattens in shape going
from low-sSFR to high-sSFR.  Our best-fit decomposition model, which is based
on a stellar-mass scaled LMXB XLF ({\it red curves\/}) and a SFR scaled HMXB
XLF ({\it blue curves\/}), is plotted in each panel.  We note that since the
y-axis is normalized by SFR, the HMXB model is the same in each of the three
panels.
}
\end{figure*}

We therefore constructed a {\it decomposition XLF model} that contained all of
the above elements.  For each of the \xray\ point sources, we identified local
estimates of the $M_\star$ and SFR from our maps, and computed the sSFR
associated with that \xray\ source location (i.e., the sSFR within a $\approx$400~pc region around the \xray\ source).  We then sorted the \xray\ point
source catalog by local sSFR and broke up the sample into 28 sSFR bins (three \xray\
sources per sSFR bin) that progressed from lowest to highest sSFR.  For each of
the sSFR bins, we computed the total areal extent across M51 that contained
sSFR values within the range defined by the bin, and estimated the number of
CXB sources expected over this area (see $\S$3.2).  Following Equation~(2), we
computed the \xray\ point source completeness function, $\xi({\rm
sSFR}_i,L_{\rm X})$, for each of the 28 sSFR bins ($i \equiv$~$1 \ldots 28$).

To construct a decomposition XLF model, we used the past results from Zhang
\etal\ (2012) and Mineo \etal\ (2012) as guidance on the basic forms of the
LMXB and HMXB XLFs, respectively.  From Zhang \etal\ (2012), the LMXB XLF
within massive ellipticals exhibits two breaks located at $\approx$$2 \times
10^{37}$~\lum\ and $\approx$$2.5 \times 10^{38}$~\lum\ in the 2--10~keV band
adopted here.  These breaks are consistent with our findings for the
galaxy-wide XLF studied in $\S$3.2 above (see also Fig~4).  We therefore
modeled the LMXB XLF as a power-law with two breaks fixed at these values, but
with a normalization that scales with stellar mass.  The two-break power law
model has the form:
\begin{equation}
\begin{split}
h(L_{\rm X}) 
 \equiv \left \{ \begin{array}{lr} L_{\rm X, 38}^{-\alpha_1}  & (L_{\rm X, 38}
\le L_{\rm b, 1}) \\ 
L_{\rm b, 1}^{\alpha_2 - \alpha_1}L_{\rm X,38}^{-\alpha_2}, & \;\;\;\;(L_{\rm
b, 1} < L_{\rm X, 38} \le L_{b, 2})\\ 
L_{\rm b, 1}^{\alpha_2 - \alpha_1} L_{\rm b, 2}^{\alpha_3 - \alpha_2}L_{\rm X,38}^{-\alpha_3}, & \;\;\;\;(L_{\rm b,
2} < L_{\rm X, 38} \le L_c)\\ 
0.  & (L_{\rm X, 38} > L_c) \\ \end{array}
  \right.
\end{split}
\end{equation}
From Mineo \etal\ (2012), galaxies with sSFR~$\simgt 10^{-10}$~yr$^{-1}$ (i.e.,
for the inferred HMXB population) have XLFs that are more consistent with a
single power-law going out to $L_{\rm X} \approx$~\hbox{(1--10)}~$\times
10^{40}$~\lum.  Thus, we chose to model the XLF as a single power-law (i.e.,
$f(L_{\rm X})$ defined in Eqn.~(1)), with a normalization that scales with SFR
and a cut-off at $10^{40}$~\lum.

In relation to the 28 sSFR bins defined above, we expect that the XLF shape
should go from looking more like a broken power law at low-sSFR to more like a
single power law at high-sSFR.  Since we have constructed each sSFR bin to
include only three \xray\ sources, and expect at least some of these sources to
be background sources, it is not easy to visually see such an effect.  However,
a broader binning of the data into three sSFR bins clearly reveals such a
trend.  In Figure~6, we show the total observed XLFs for three sSFR bins,
normalized by the SFR appropriate for each bin (i.e., the total SFR across the galactic extent from regions within a given sSFR bin).  As expected, we
indeed see the most obvious indication of a break in the lowest-sSFR bin, a
trend that has been indirectly observed since early \chandra\ XLF studies of various
nearby galaxies and subgalactic regions (e.g., bulges, spiral arms, and
interarm regions; e.g., Kilgard \etal\ 2002, 2005; Trudolyubov \etal\ 2002;
Kong \etal\ 2003; Soria \& Wu~2003).  We also see that the SFR-normalized XLF
reaches the largest number of \xray\ sources per SFR in the lowest-sSFR bin.
If all \xray\ sources throughout the galaxy were HMXBs, we would expect the
SFR-normalized XLF to be roughly the same in each panel.  The increased
normalization towards low-sSFR is a direct indication of a corresponding
increase in the number of LMXBs per unit SFR for low sSFR.

\begin{table}
\begin{center}
\caption{Summary of Fits to X-ray Luminosity Functions}
\begin{tabular}{llll}
\hline\hline
 \multicolumn{1}{c}{Model Description} &  \multicolumn{1}{c}{Parameter} & \multicolumn{1}{c}{Param Value} & \multicolumn{1}{c}{Units} \\
\hline
\multicolumn{4}{c}{Galaxy-Wide Power-law Models}\\
\hline
single power law \dotfill & $K_s$ & 9.3$^{+1.2}_{-1.4}$ & \\
                             & $\gamma$ & 1.49$^{+0.06}_{-0.08}$ & \\
        & {\ttfamily cstat} & 194.0 &  \\
        & $P(\ge${\ttfamily cstat}$)$ & 0.75 &  \\
\\
broken power law \ldots\ldots\ldots\ldots & $K_b$ & 11.2$^{+1.4}_{-2.4}$ &  \\
                             & $\alpha_1$ & 1.38$^{+0.09}_{-0.10}$ & \\
                             & $\alpha_2$ & 1.94$^{+0.38}_{-0.34}$ & \\
                             & $L_b$ & 2.80$^{+1.59}_{-0.39}$ & $10^{38}$~\lum \\
        & {\ttfamily cstat} & 191.5 &  \\
        & $P(\ge${\ttfamily cstat}$)$ & 0.66 &  \\
\hline
\multicolumn{4}{c}{LMXB and HMXB Decomposition Model}\\
\hline
LMXB component \dotfill & $K_{\rm LMXB}$ & 27.89$^{+67.33}_{-10.04}$ &  ($10^{11}$~\msol)$^{-1}$ \\
                             & $\alpha_1$ & 1.44$^{+0.03}_{-0.62}$&  \\
                             & $L_{b,1}$ & 0.2$^\dagger$ & $10^{38}$~\lum \\
                             & $\alpha_2$ & 1.32$^{+0.28}_{-0.14}$ & \\
                             & $L_{b,2}$ & 2.5$^\dagger$ & $10^{38}$~\lum \\
                             & $\alpha_3$ &  3.0$^\dagger$ & \\
HMXB component \dotfill & $K_{\rm HMXB}$ & 1.10$^{+0.83}_{-0.71}$ & (\msol~yr$^{-1}$)$^{-1}$ \\
                             & $\gamma$ & 1.24$^{+0.25}_{-0.23}$ &  \\
        & $C_{\rm global}$ & 710.3 &  \\
        & $P(\ge C_{\rm global}$) & 0.13 &  \\
\hline
\multicolumn{4}{c}{Star-Formation History Model}\\
\hline
        & $K_0$ & 52$^{+24}_{-15}$ & ($10^{11}$~\msol)$^{-1}$  \\
        & $\kappa$ & 0.51$^{+0.42}_{-0.27}$ & \\
        & $\alpha_1$ & 1.42$^{+0.06}_{-0.21}$ &  \\
        & $L_{b,1}$ & 0.2$^\dagger$ & $10^{38}$~\lum  \\
        & $\alpha_2({\rm 10~Gyr})$ & 1.37$^{+2.20}_{-0.46}$ & \\
        & $L_{b,2}$ & 2.5$^\dagger$ & $10^{38}$~\lum  \\
        & $\alpha_3({\rm 10~Gyr})$ &  3.0$^\dagger$ & \\
        & $C_{\rm global}$ & 748.5 &  \\
        & $P(\ge C_{\rm global}$) & 0.10 &  \\
\hline
\end{tabular}
\end{center}
$^\dagger$Indicates parameter was fixed in fitting procedure (see text for details).
\end{table}

%
%
\begin{figure}
\figurenum{7}
\centerline{
\includegraphics[width=9cm]{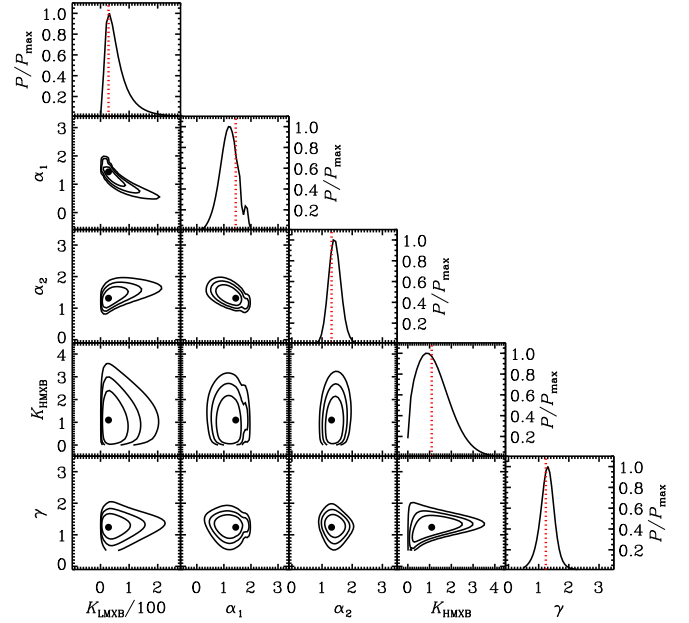}
}
\vspace{0.1in}
\caption{
Marginalized best-fit parameter estimate contours for pairs of parameters ({\it
contour plots} with 68\%, 87\%, and 95\% confidence contours drawn) and
probability density distributions for single parameters ({\it continous
curves\/}), pertaining to the decomposition model described in $\S$3.3.  The
global best-fit parameter values are indicated in each of the contour plots
with a single black filled circle and vertical red dotted lines in the
probability density distribution diagrams.  All parameters in our model are
well constrained, albeit with large fractional uncertainties for some
parameters.
}
\end{figure}

%
%
\begin{figure*}
\figurenum{8}
\centerline{
\includegraphics[width=9cm]{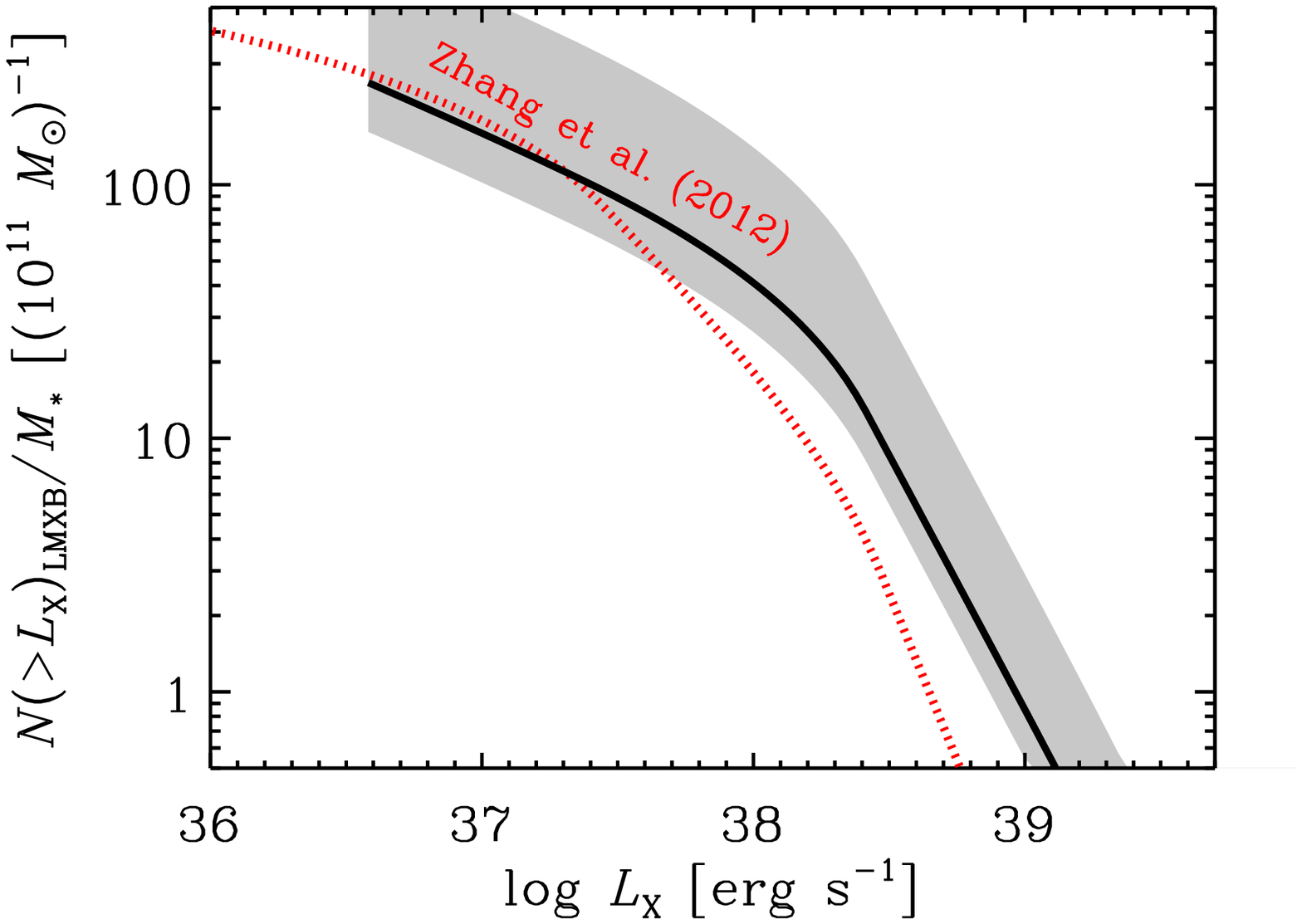}
\hfill
\includegraphics[width=9cm]{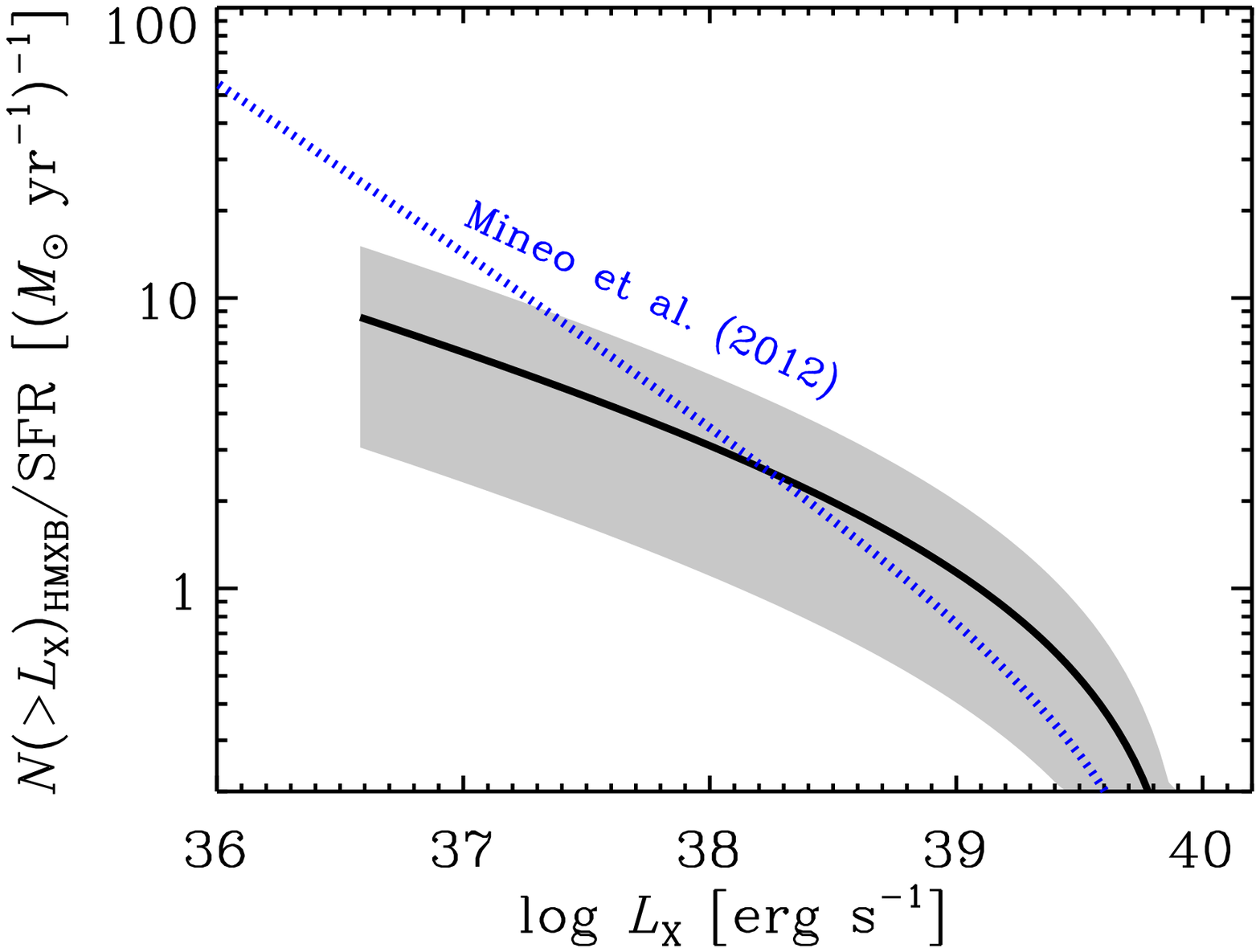}
}
\caption{
Our best-fit stellar-mass normalized LMXB XLF model ($a$) and SFR normalized
HMXB XLF model ($b$) ({\it solid black curves\/}).  For comparison, we have
overlayed, with dotted curves, the elliptical-galaxy based LMXB XLF derived
by Zhang \etal\ (2012) and the high-sSFR late-type galaxy based HMXB XLF
derived by Mineo \etal\ (2012).  Our recovered LMXB and HMXB XLFs for M51
appear broadly consistent with those from the literature with some key
differences, likely due to differences in GC populations, SFHs, metallicities, and assumptions between analyses (see detailed
discussion in $\S$3.3).
}
\end{figure*}

Using the full set of 28 sSFR bins, for the $i$th bin we constructed the
following scaled model for the overall observed XLF:
\begin{equation}
\begin{split}
dN/dL_{\rm X}({\rm sSFR}_i) = \xi({\rm sSFR}_i, L_{\rm X}) [{\rm SFR}_i K_{\rm
HMXB} f(L_{\rm X}) + \\
M_{\star,i} K_{\rm LMXB} h(L_{\rm X}) + dN/dL_{\rm X}({\rm sSFR}_i,{\rm CXB})],
\end{split}
\end{equation}
where $f(L_{\rm X})$ and $h(L_{\rm X})$ are the un-normalized single and
double-break broken power-law functions (see Eqn.~(1) and (5)) and $K_{\rm
LMXB}$ and $K_{\rm HMXB}$ are the corresponding $M_\star$ and SFR scaled XLF
normalizations, respectively.  This model contains five parameters: $K_{\rm
HMXB}$ and $\gamma$ for the SFR-scaled HMXB XLF and $K_{\rm LMXB}$, $\alpha_1$,
and $\alpha_2$ for the $M_\star$-scaled LMXB XLF.  Since $\alpha_3$ describes
the $L_{\rm X} \simgt 2.5 \times 10^{38}$~\lum\ slope, and few sources are
present, we could not constrain its value.  We chose to fix its value at
$\alpha_3 = 3.0$, a value consistent with field LMXBs in elliptical galaxies
(see extended discussion in $\S$3.3 and Peacock \etal\ 2017 for motivation).

To determine best-fit parameters, we made use of a summed {\ttfamily cstat}
value to obtain a global statistic, $C_{\rm global}$, following
\begin{equation}
C_{\rm global} = \sum_{i=1}^{28} C({\rm sSFR}_i),
\end{equation}
where $C({\rm sSFR}_i)$ is the {\ttfamily cstat} value for the $i^{\rm th}$
sSFR bin.  Our model was thus fit by minimizing $C_{\rm global}$ following the
procedure that we developed in $\S$3.2.  The best fit parameters for our LMXB
and HMXB decomposition model are summarized in Table~2. 

In Figure~6, we show the SFR-normalized best-fit model XLFs (in cumulative
form) for the three sSFR bins discussed above with HMXB (blue curves) and LMXB
(red curves) contributions indicated.  By construction, our SFR-normalized HMXB
XLF model is the same in each of the panels of Figure~6; however, the
contribution from the LMXB XLF model grows with decreasing sSFR.  
This simple model provides a reasonable characterization
of the basic scaling of the XLFs in all three panels.  We re-iterate that our model is a single
model that contains an HMXB XLF with normalization $K_{\rm HMXB}$ that scaled
linearly with SFR and an LMXB component with normalization $K_{\rm LMXB}$ that
scales linearly with $M_\star$.  Only SFR and $M_\star$ vary between each of the
three panels in Figure~6.

We performed goodness-of-fit simulations, as described in $\S$3.2, and found
$P(\ge C_{\rm global}) = 0.13$, suggesting that the data are marginally
consistent with the adopted model.  In Figure~7, we show the marginalized
probability density functions and contours for parameter pairs.  All parameters
in the model are constrained by the data, albeit with large fractional
uncertainties for some of the parameters.  For example, the normalization terms
are not well constrained, primarily due to the correlation between the LMXB
normalization, $K_{\rm LMXB}$, and its slopes, $\alpha_1$ and $\alpha_2$.

In Figure~8, we show the decomposed LMXB and HMXB XLFs normalized by $M_\star$
and SFR, respectively, and display the best models derived by Zhang \etal\
(2012; LMXBs) and Mineo \etal\ (2012; HMXBs), for comparison.  We find
reasonable agreement between our derived LMXB and HXMB XLFs in M51 and those
found for large populations of galaxies, with some differences.  The equivalent
elliptical galaxy LMXB XLF slopes from Zhang \etal\ (2012) are $\alpha_1 =
1.02^{+0.07}_{-0.08}$, $\alpha_2 = 2.06^{+0.06}_{-0.05}$, and $\alpha_3 =
3.63^{+0.67}_{-0.49}$.  Our LMXB XLF shows a somewhat shallower value of
$\alpha_2 = 1.32^{+0.28}_{-0.14}$, but is otherwise consistent with the slopes
from Zhang \etal\ (2012).  For the HMXBs, we find a marginally shallower XLF
slope of $\gamma = 1.24^{+0.25}_{-0.23}$ compared with the $\gamma = 1.58 \pm
0.02$ value obtained by Mineo \etal\ (2012) for high-sSFR galaxies; however,
the HMXB XLFs appear to be consistent at least for $L_{\rm X} \simgt
3 \times 10^{37}$~\lum, where the constraints are best.

The near consistency between our recovered LMXB and HMXB XLFs with those in the
literature is encouraging, and there are a number of factors that could explain
any residual differences.  For example, the LMXB XLF derived by Zhang \etal\
(2012) was based primarily on massive elliptical galaxies, which have a larger
number of globular clusters (GCs) per unit stellar mass than late-type galaxies
like M51. Due to their high stellar densities and enhanced stellar interaction
rates over stellar systems in the galactic field, GCs contain significant
numbers of LMXBs that form via dynamical interactions (see, e.g., Benacquista
\& Downing~2013 for a review).  As such, we would expect there to be more GC
LMXBs per unit galactic stellar mass for ellipticals over M51, and thus an
elevated LMXB XLF from the Zhang \etal\ (2012) study.  However, there are also
indications that the stellar-mass normalized {\it field} LMXB XLF is larger for
younger stellar populations (e.g., Kim \etal\ 2010; Lehmer \etal\ 2014), which
would presumably favor an enhanced stellar-mass normalized field LMXB XLF in
M51 over that of typical ellipticals, which have older SFHs.  It is possible
that the combination of these effects has led to similar LMXB XLFs for M51 and
typical ellipticals.

%
%
\begin{figure}
\figurenum{9}
\centerline{
\includegraphics[width=9cm]{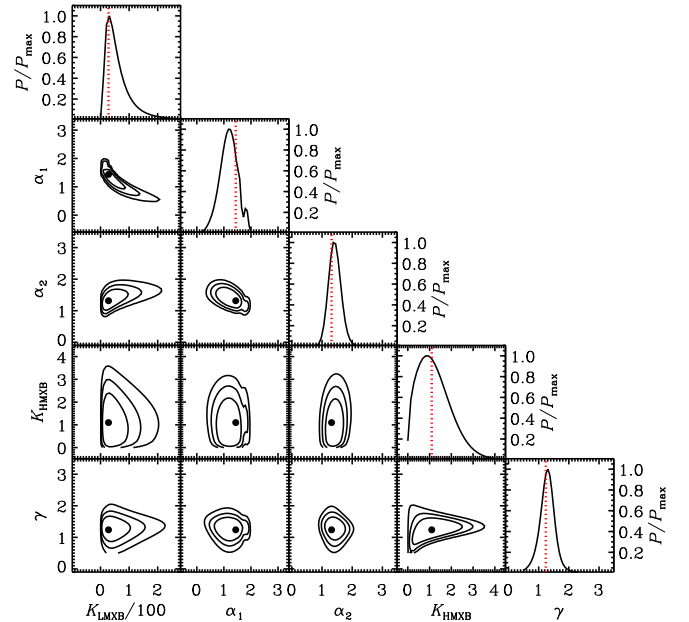}
}
\caption{
Galaxy-wide cumulative XLF for M51 ({\it gray circles with error bars\/}; same
as Fig.~4$b$), with our best LMXB ({\it red curve\/}) and HMXB ({\it blue
curve\/}) decomposition model overlayed.  As per Fig.~4, the dotted curve represents the estimated CXB contribution.
The decomposition model was
constructed by multiplying the galaxy-wide stellar mass and SFR by the
respective LMXB and HMXB model terms provided in equation~(6).  We infer that
HMXBs dominate the cumulative XLF for M51 above $L_{\rm X}
\approx$~\hbox{(3--5)}~$\times 10^{38}$~\lum, with LMXBs dominating at lower
luminosities.  The integrated XLFs suggest that LMXBs and HMXBs provide similar
contributions to the overall integrated \xray\ power output of the galaxy.
}
\end{figure}

The HMXB XLF from Mineo \etal\ (2012) was constructed for a sample of galaxies
with sSFR~$\simgt 10^{-10}$~yr$^{-1}$ and in some galaxies, like M51, the bulge
regions were excluded in an effort to isolate the HMXB population.  In their
analyses, however, they assumed that the LMXB contributions in the disk regions
was negligible and did not make any corrections for this population.  In our
analysis, we found that our best models predict at least some contribution from
LMXBs even in the highest-sSFR regions.  In the far-right panel of Figure~6, we
show the decomposed XLF for the highest sSFR regions in M51.  Our best model
suggests that the steeper LMXB component of that model is comparable to the
HMXBs at $L_{\rm X} \simlt 10^{38}$~\lum, and taken together, the LMXB plus
HMXB XLF takes on a steeper slope that is more consistent with the Mineo \etal\
(2012) XLF.  It is therefore plausible that the true HMXB XLF is flatter than
previously reported (as seen in the right panel of Fig.~8), and past
investigations may not be accounting for an important LMXB contribution at low
$L_{\rm X}$.  However, further studies of additional galaxies would be needed
to verify this claim.  M51 has a unique SFH and history of abundance enrichment
that may not be representative of galaxies as a whole.  These properties (i.e.,
SFH and metallicity) are known to influence the formation of XRBs (see, e.g.,
Fragos \etal\ 2013).

%
%
\begin{figure*}
\figurenum{10}
\centerline{
\includegraphics[width=17cm]{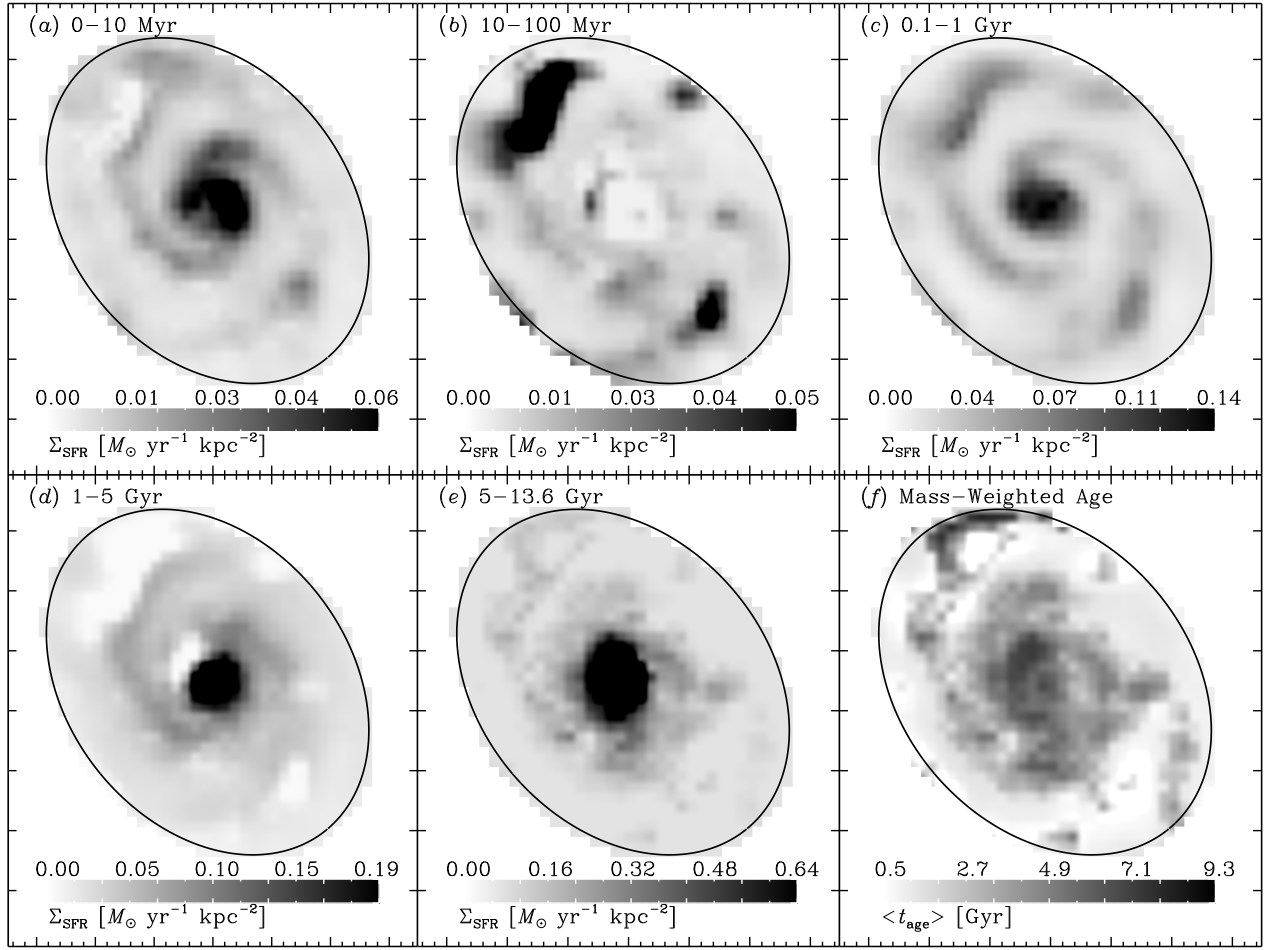}
}
\vspace{0.1in}
\caption{
SFH maps of M51 in five stellar-age bins: 0--10~Myr, 10--100~Myr, 0.1--1~Gyr,
1--5~Gyr, and 5--13.6~Gyr (see annotations).  The scales of these images are
linear and vary between frames to highlight the spatial distributions of the
star-formation activity within each stellar-age range.  The bottom-right frame
shows the spatial distribution of the mass-weighted mean stellar age, $\langle t_{\rm age} \rangle$,
a quantity we use to
separate regions of different SFHs (see $\S$3.4 for details).
}
\end{figure*}

%
%
\begin{figure}
\figurenum{11}
\centerline{
\includegraphics[width=9cm]{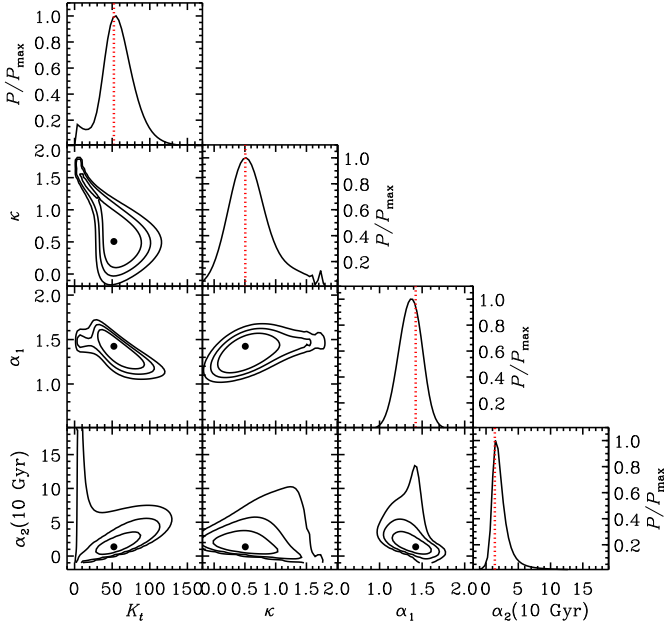}
}
\vspace{0.1in}
\caption{
Same as Figure~7, but for parameters involved in the SFH XLF decomposition model described in $\S$3.4.
}
\end{figure}

%
%
\begin{figure*}
\figurenum{12}
\centerline{
\includegraphics[width=6cm]{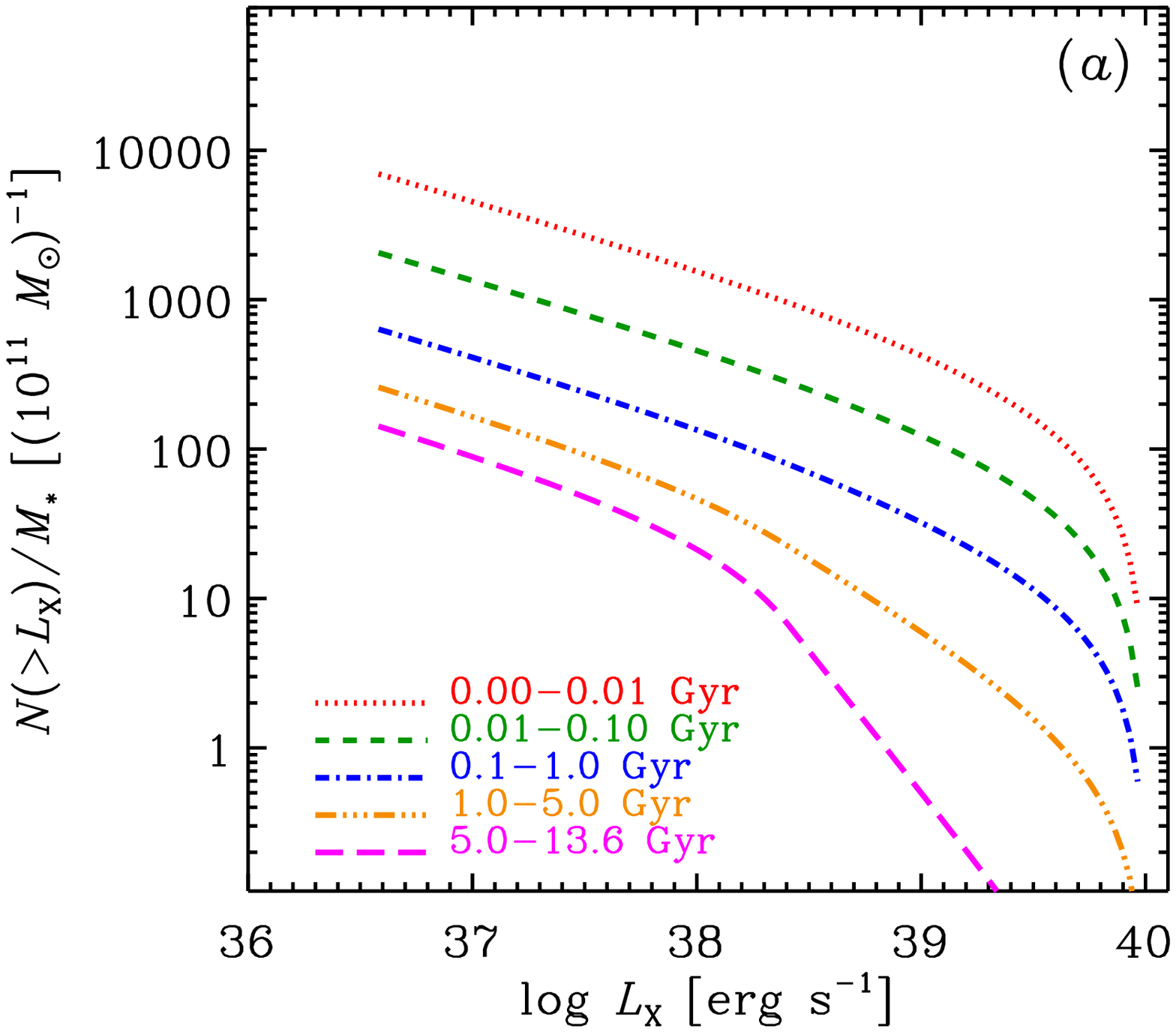}
\hfill
\includegraphics[width=6cm]{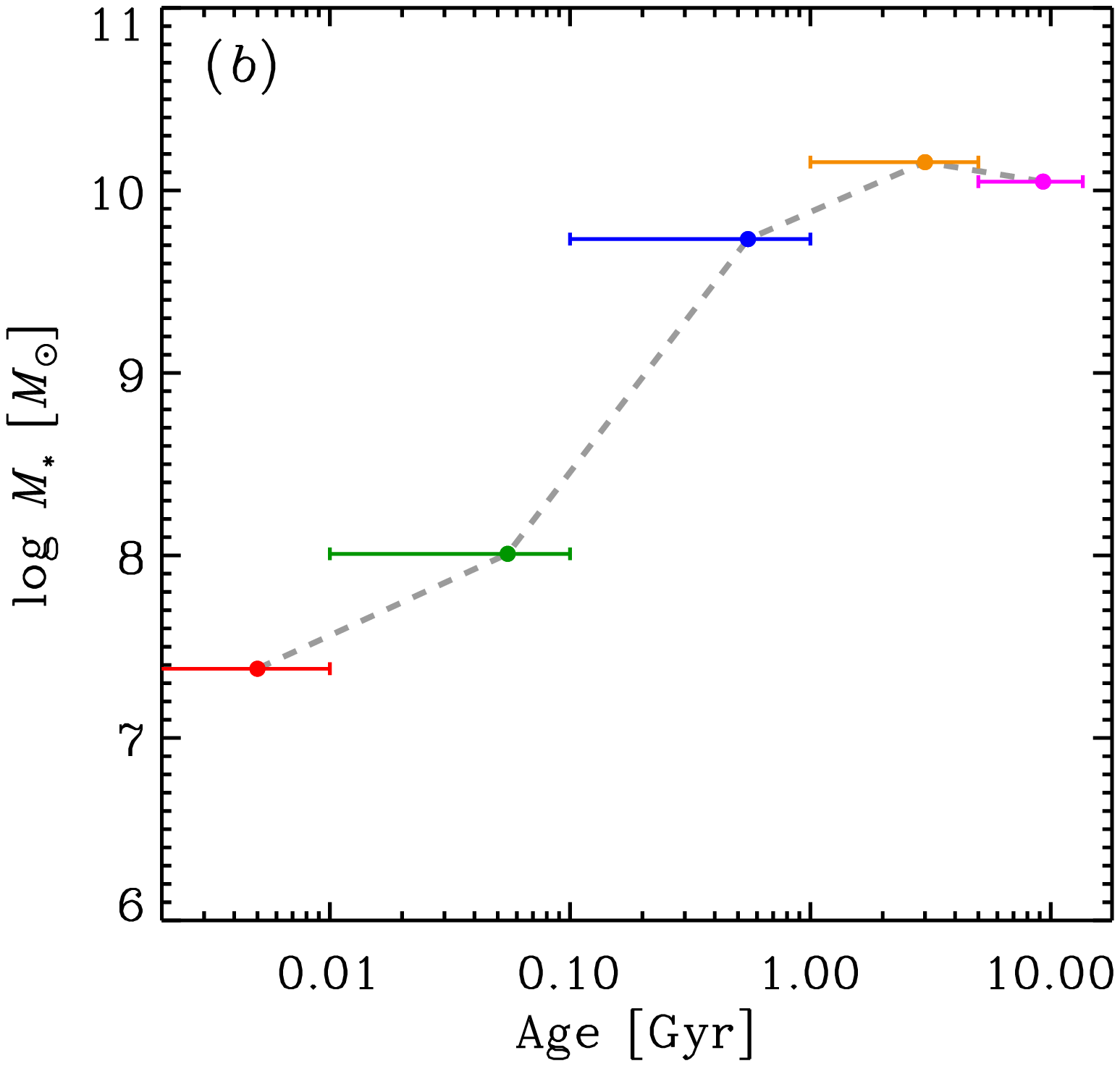}
\hfill
\includegraphics[width=6cm]{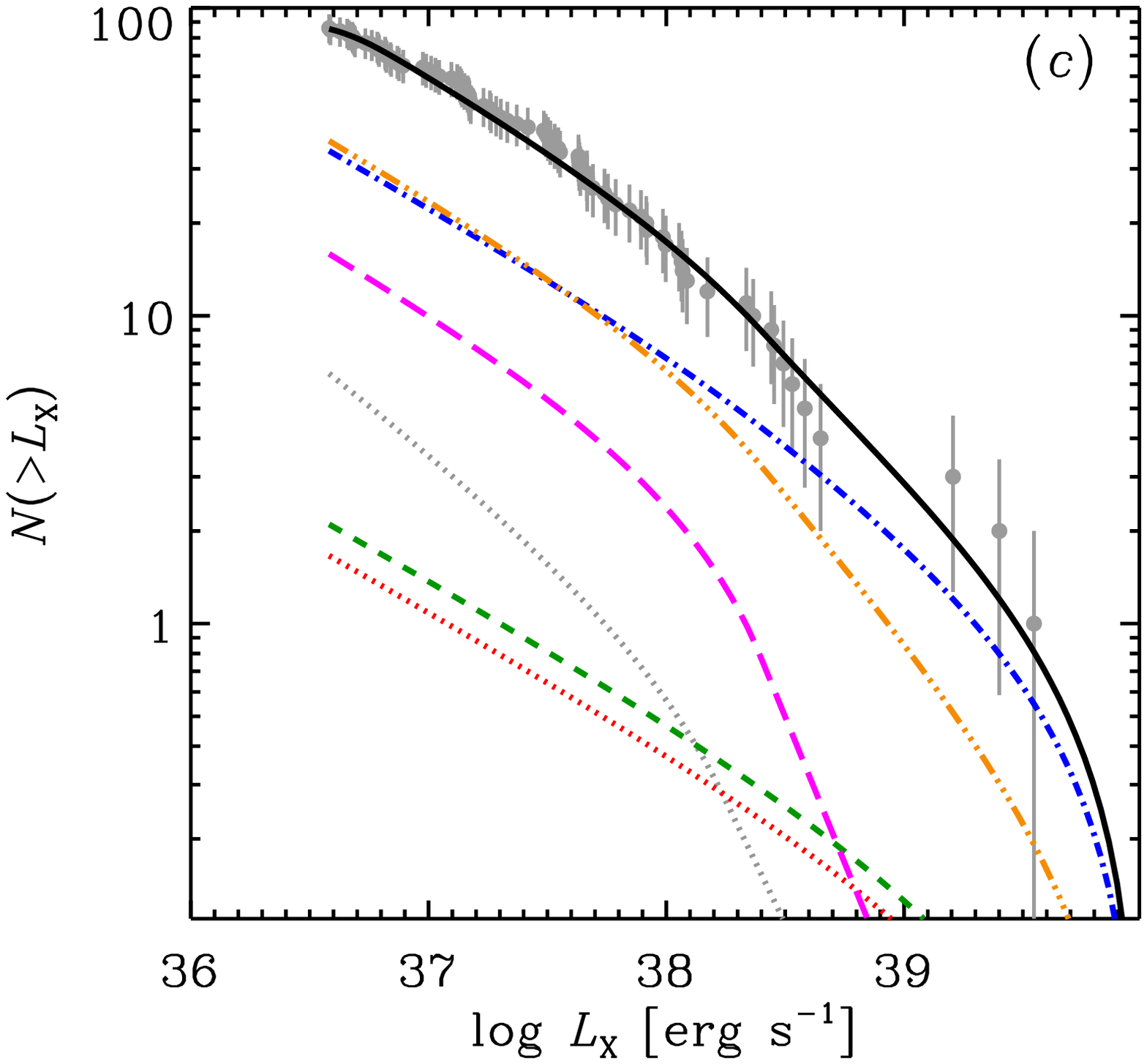}
}
\caption{
($a$) Best-fit SFH XLF decomposition model showing how the
XRB XLF evolves with stellar age.  Each curve shows the XLF for a population of
XRBs within the designated age range, normalized by the stellar mass of that
same age range.  Our model indicates that the stellar-mass normalized XRB XLF
declines in normalization by $\sim$3--3.5 orders of magnitude and the overall XLF
slopes steepen significantly between $\sim$0--10~Myr and $\sim$5--13.6~Gyr.
($b$) Integrated galaxy-wide SFH for M51, expressed in terms of the current
stellar mass contributions from the five stellar age bins.  The bins have been
color coded to match those defined in panel~$a$.  The majority of the stellar
mass in M51 is attributed to the 1--5~Gyr population. ($c$) Galaxy-wide
cumulative XLF for M51 ({\it gray circles with error bars}; same as Fig.~4$b$).  As per Fig.~4, the dotted curve represents the estimated CXB contribution.
Our SFH XLF model (panel $a$), folded in with the SFH of M51 (panel~$b$), is
shown as a black curve with contributions from each of the five stellar age
bins indicated with colored curves, which have the same meaning as they did in
panel~$a$.  Our SFH model provides an excellent characterization of the
galaxy-wide XLF of M51 and contains the same number of fitting parameters as
the broken power-law model.
}
\end{figure*}

%
%
\begin{figure*}
\figurenum{13}
\centerline{
\includegraphics[width=14cm]{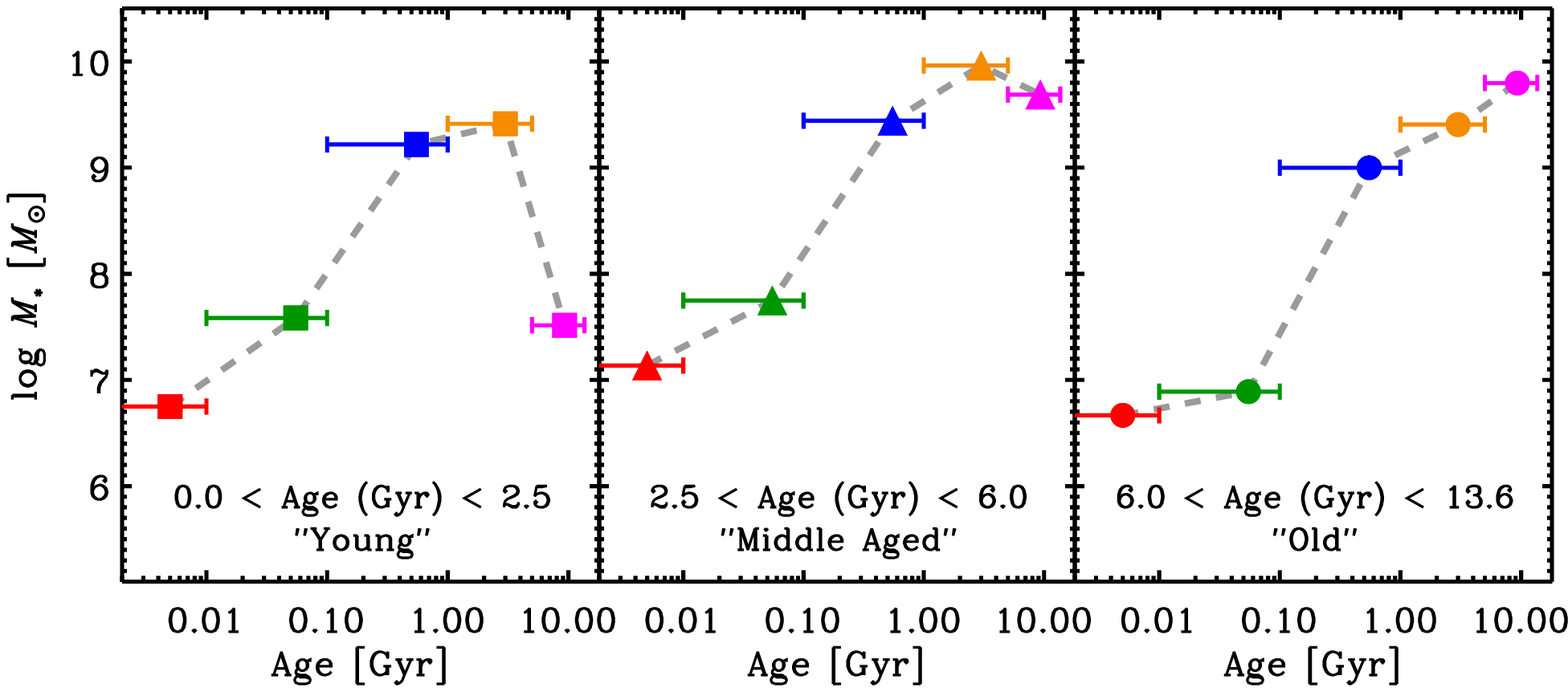}
}
\vspace{0.2in}
\centerline{
\includegraphics[width=14cm]{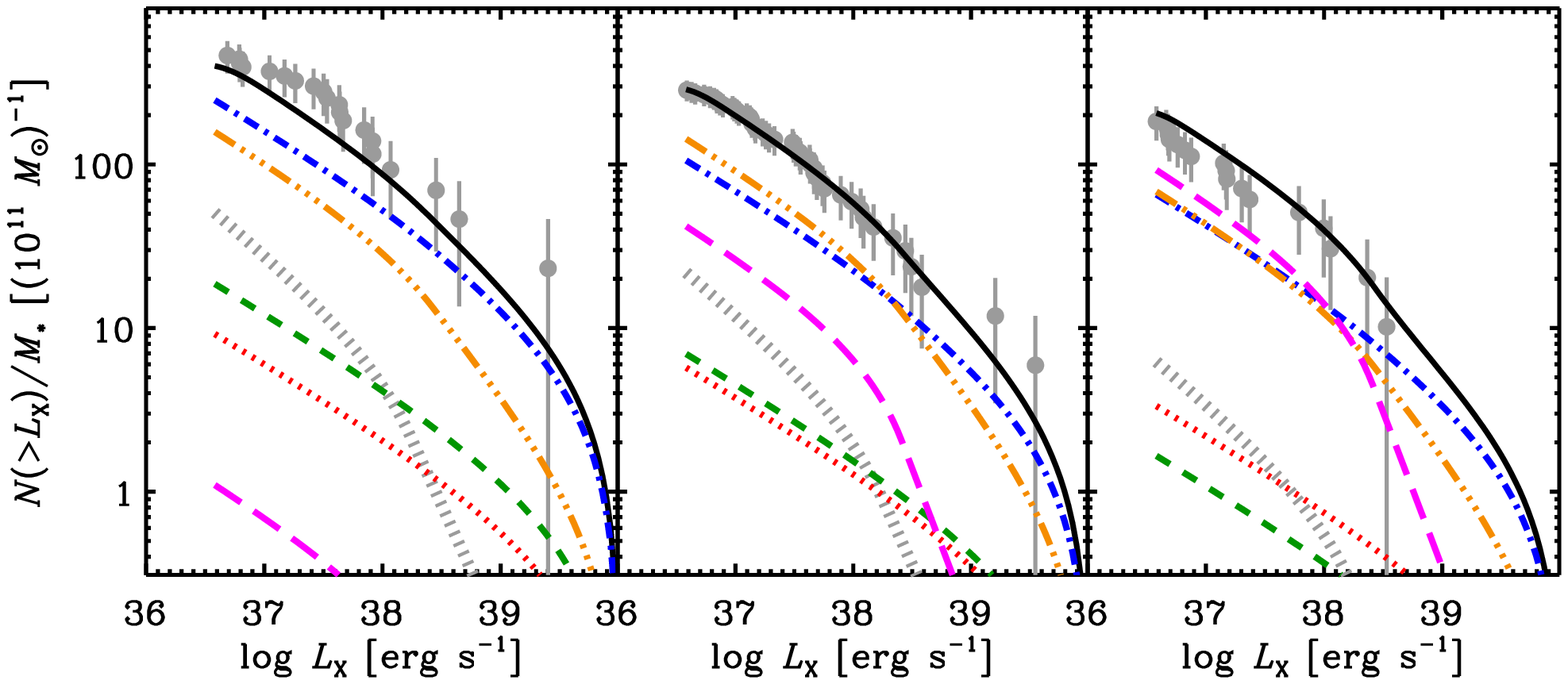}
}
\caption{
({\it Top Panels\/}) Five-step SFHs for three subgalactic regions that were
selected by mass-weighted mean stellar ages, with the ranges annotated in the
lower portion of each panel.  Here we divided M51 into ``young'' ($\langle
t_{\rm age} \rangle =$~0--2.5~Gyr; {\it left panel\/}), ``middle aged''
($\langle t_{\rm age} \rangle =$~2.5--6~Gyr; {\it middle panel\/}), and
``old'' ($\langle t_{\rm age} \rangle =$~6--13.6~Gyr; {\it right panel\/})
populations.  
({\it Bottom Panels\/}) Observed stellar-mass normalized XLFs for the three
subgalactic regions ({\it gray points with error bars\/}).  Similar to
Figure~4, the expected CXB contributions have been shown as gray dotted curves.
The five stellar-age contributions, estimated from our best XRB evolution
model, have been shown with colored curves.  The color of each curve in the
bottom panel provides the estimated contribution to the XLF from populations
within the stellar age bin of the same color displayed in the top panel.  These
include populations from 0--10~Myr ({\it red\/}), 10--100~Myr ({\it green\/}),
0.1--1~Gyr ({\it blue\/}), 1--5~Gyr ({\it orange\/}), and 5--13.6~Gyr ({\it
magenta\/}).  The total predicted XLF for each subgalactic region is displayed
as a solid black curve.
}
\vspace{0.1in}
\end{figure*}

In Figure~9, we show the XLF for the entire galaxy-wide \xray\ detected
point-source population and the resulting best-fit decomposition model
appropriate for our estimates of SFR~=~2.0~\sfr\ and $M_\star$~=~$3.0 \times
10^{10}$~\msol.  The best-fit decomposition model provides a very good
characterization of the galaxy-wide XLF.  

\subsection{The Star-Formation History XLF Decomposition}

The above decomposition of the point-source XLF of M51 into its LMXB and HMXB
contributions provides a first-order assessment of how the XLF of XRB
populations changes as stellar populations age.  To first order, the XRB XLF
within young stellar populations can be characterized as having a constant
shallow power-law slope, extending to high \xray\ luminosities, while the XRB XLF for old
stellar populations has a steeper overall slope with two well-documented breaks.
To date, there have been very few empirical studies quantifying
how the XRB XLF shape transitions with XRB popuation age from the relatively flat HMXB XLF
to the steeper LMXB XLF.  There has been some evidence that over the 2--10~Gyr
timescale, the XLF of field LMXBs in elliptical galaxies does indeed become
steeper with increasing age (see, e.g., Kim \& Fabbiano~2010; Lehmer \etal\ 2014) and theoretical
XRB population synthesis models find similar behaviors (e.g., Belczynski \etal\
2004; Fragos \etal\ 2008; Tzanavaris \etal\ 2013); however, the details of how
the XLF shape changes throughout the transition remain highly uncertain.  
In this section, we apply a new empirical method for estimating how the XRB
XLF evolves with time based on M51 data alone.  

Similar to our LMXB and HMXB XLF decomposition procedure, discussed in $\S$3.3,
we can decompose the general XRB XLF into contributions from stellar
populations that span the full SFH of the galaxy.  In Eufrasio \etal\ (2017),
we constructed spatially-resolved SFH maps of M51, which contain five maps
where each pixel contains the contributions to the stellar mass 0--10~Myr,
10--100~Myr, 0.1--1~Gyr, 1--5~Gyr, and 5--13.6~Gyr old populations; these maps
are displayed in Figure~10.  
In order to distinguish between regions that have strong contributions from
``young'' stellar populations versus ``old'' populations, we made use of the
mass-weighted stellar age, which is calculated for the $i$th pixel following
\begin{equation}
\langle t_{{\rm age},i} \rangle = \frac{\sum_{j=1}^5 m_{\star,i,j}
t_j}{M_{\star,i}}
\end{equation}
where $m_{\star,i,j}$ and $t_j$ are the stellar mass contributions and
bin-central stellar ages (i.e., $\{t\} =$~5~Myr, 55~Myr, 550~Myr, 3~Gyr, and
9.3~Gyr) for the $i$th pixel and $j$th SFH bin, and $M_{i, \star}$
is the total stellar mass in the $i$th pixel.  In the bottom-right panel of
Figure~10, we provide a spatial map of $\langle t_{\rm age} \rangle$.  

Following a similar approach to that in $\S$3.3, we determined the value of
$\langle t_{\rm age} \rangle$ associated with populations in the vicinity of
each \xray\ point source, sorted our \xray\ point source catalog by $\langle
t_{\rm age} \rangle$, and binned the \xray\ point source sample into 28 bins of
$\langle t_{\rm age} \rangle$ with three \xray\ sources per bin.  For each
$\langle t_{\rm age} \rangle$ bin, we calculated the area, CXB
contributions ($dN/dL_{\rm X}(\langle t_{\rm age} \rangle,{\rm CXB})$), and
\xray\ source completeness functions ($\xi(\langle t_{\rm age} \rangle, L_{\rm
X})$).

Next, we constructed an XRB XLF model that evolves with age.  
Given our constraints
on HMXB and LMXB XLFs, as well as those in the literature, we expect that the
generalized XRB XLF shape evolves from a single power-law shape at timescales
around $\sim$10--100~Myr (i.e., for HMXBs) to a broken power-law shape on
$\sim$1--10~Gyr timescales.  Over these broad timescales, the XLF normalization
(i.e., number of XRBs per unit stellar mass) is also expected to decline
(see, e.g., Mineo \etal\ 2012).  We incorporate these basic
behaviors into a generalized model of the XRB XLF evolution with age.  The XLF model for the $i$th $\langle t_{\rm age} \rangle$ bin is defined as follows:
\begin{equation}
\begin{split}
dN/dL_{\rm X}(\langle t_{{\rm age}, i} \rangle) =  \xi(\langle t_{{\rm age}, i} \rangle, L_{\rm X}) \times \\
\left \{ \left[\sum_{j=1}^5 m_{\star,i,j} h(L_{\rm X},t_j)
\right] + dN/dL_{\rm X}(\langle t_{{\rm age}, i} \rangle,{\rm CXB}) \right \},
\end{split}
\end{equation}
where the $j$-term summation is a summation over contributions from the five
stellar-age bins defined above. As such, $m_{\star,i,j}$ is the stellar mass of
the $i$th $\langle t_{\rm age} \rangle$ bin and $j$th stellar-age bin in the SFH.  The term
$h(L_{\rm X},t_j)$ provides the XRB XLF model contribution from the $j$th
stellar age bin, which is an evolving broken power-law model defined as:
\begin{equation}
\begin{split}
h(L_{\rm X},t) 
 \equiv K(t) \left \{ \begin{array}{lr} L_{\rm X,
38}^{-\alpha_1}  & (L_{\rm X, 38} \le L_{\rm b, 1}) \\ 
L_{\rm b, 1}^{\alpha_2(t) - \alpha_1}L_{\rm X,38}^{-\alpha_2(t)},
& \;(L_{\rm b, 1} < L_{\rm X, 38} \le L_{b, 2})\\ 
L_{\rm b, 1}^{\alpha_2(t) - \alpha_1} L_{\rm b, 2}^{\alpha_3(t) - \alpha_2(t)}L_{\rm X,38}^{-\alpha_3(t)},
& \;(L_{\rm b, 2} < L_{\rm X, 38} \le L_c)\\ 
0,  & (L_{\rm X, 38} > L_c) \\ \end{array}
  \right.
\end{split}
\end{equation}
\begin{equation}
K(t) = K_0 \left( \frac{t}{\rm 1~Gyr} \right)^{-\kappa},
\end{equation}
\begin{equation}
\alpha_2(t) = \alpha_1 \left [1 + (\alpha_{2}({\rm 10~Gyr})/\alpha_1 - 1) \left(\frac{t}{{\rm 10~Gyr}} \right) \right ],
\end{equation}
and
\begin{equation}
\alpha_3(t) = \alpha_1 \left [1 + (\alpha_{3}({\rm 10~Gyr})/\alpha_1 - 1) \left(\frac{t}{{\rm 10~Gyr}} \right) \right ].
\end{equation}

The above model is a broken power-law, of the same form as that given in
equation~(5), but contains age-variable normalization and slopes.  The
parameterizations of the time-variable components $K(t)$, $\alpha_2(t)$, and
$\alpha_3(t)$ were chosen to mimic a transition from an HMXB-like XLF at $t=0$
and a LMXB-like XLF observed for elliptical galaxies with $t \approx
10$~Gyr.  By construction, the model starts off as a single power-law sloped
XLF at $t=0$ and allows the shape to transition to a broken power-law form (if
needed by the data) at 10~Gyr with breaks located at the well-known breaks
$L_{b,1} = 0.2$ and $L_{b,2} = 2.0$, values that we fix in our model.  We
performed fitting to obtain values for four of the parameters: $K_0$, $\kappa$,
$\alpha_1$, and $\alpha_2({\rm 10~Gyr})$, which collectively constrain how the
normalization and shape of the XRB XLF varies as a function of age.  

In principle, the bright-end XLF slope at 10~Gyr, $\alpha_3({\rm 10~Gyr})$,
could have also been used in the fitting process; however, we find that our
data provide only a very weak constraints of $\alpha_3({\rm 10~Gyr}) \simgt 2$.
Furthermore, we have several constraints on $\alpha_3({\rm 10~Gyr})$ already
from studies of elliptical galaxies.  For example, Zhang \etal\ (2012)
determined $\alpha_3({\rm 6~Gyr}) = 3.63^{+0.67}_{-0.49}$ based on the
collective XLF of 20 massive nearby elliptical galaxies, which is consistent
with $\alpha_3({\rm 10~Gyr}) \approx 5$ based on our best model value of
$\alpha_1 \approx 1.4$ (see below).  Lehmer \etal\ (2014) find $\alpha_3({\rm
9~Gyr}) = 3.3 \pm 1.1$ for the field LMXB population in the galaxy NGC~3379,
putting $\alpha_3({\rm 10~Gyr})$ in the range of \hbox{$\approx$2.5--5}.  For
the elliptical galaxy NGC~3115 ($t_{\rm age} \approx 9$~Gyr) Lin \etal\ (2015)
estimate $\alpha_3({\rm 9~Gyr}) \simgt 7$ ($\alpha_3({\rm 10~Gyr}) \simgt
7.7$).  We note, however, that small number statistics can dramatically affect
the measurements of $\alpha_3({\rm 10~Gyr})$ for individual galaxies.
Recently, Peacock \etal\ (2017) utilized \hst\ and \chandra\ observations of
nine nearby galaxies with ages spanning $\approx$9--11.5~Gyr to study the XLFs
of field LMXBs (see also Peacock \& Zepf~2016).  They found a value of
$\alpha_3({\rm 10~Gyr}) \approx 3$ describes well the ensemble field LMXB XLF
above $L_{b,2}$.  As such, we chose to adopt $\alpha_3({\rm 10~Gyr}) = 3$,
which is broadly consistent with all the observations.  

Using the above SFH XLF model, we determined best-fit parameters following the
same procedure developed in $\S$3.3, in which we consider the model described
in Equation~(9) globally, by summing the {\ttfamily cstat} values of all 28
$\langle t_{\rm age} \rangle$ bins and minimizing this global value (see, e.g.,
Eqn~(7)).  Our best-fit model parameters are tabulated in Table~2, and in
Figure~11, we show the marginalized probability distributions and probability
contours for parameter pairs.  We find that this model provides an equivalent fit
to the data compared to our LMXB and HMXB decomposition model, presented in $\S$3.3,
with a goodness-of-fit statistic $P(\ge C_{\rm global}) = 0.10$.  

In Figure~12$a$, we show the best-fit model as the stellar-mass normalized XRB
XLF in each of the five stellar-age bins.  It is clear that our best-fit
solution confirms the behavior that we initially predicted -- i.e., the
normalization declines and the high-$L_{\rm X}$ slope steepens with increasing
stellar age.  We note that this behavior is not required by our choice of
model.  For example, if $\kappa$ were determined to be negative, then the
normalization would increase with age, and if $\alpha_1$ were greater than or
equal to $\alpha_{3}({\rm 10~Gyr})$, then the slopes could be flat or become
shallower with increasing age.  In Figure~12$b$ we show the galaxy-wide SFH for
M51, expressed in terms of the current stellar mass contributions from each
stellar-age bin.  We can obtain an estimate of the galaxy-wide XLF by taking
each of the five components in the XLF model shown in Figure~12$a$, multiplying
them by their corresponding stellar mass from Figure~12$b$, and summing all
five contributing curves, plus the CXB contribution.  Our estimates of the
total galaxy-wide XLF, and the individual contributions from each of the five
stellar-age bins, are shown in Figure~12$c$.  Our best-fit solution predicts
that the majority of the XRBs in M51 originate from the $\approx$100~Myr to
1~Gyr population with additional contributions at $L_{\rm X} \simlt
10^{38}$~\lum\ from the other populations.

For illustrative purposes, we created Figure~13, which displays the SFHs (in
units of stellar mass per stellar-age bin) and stellar-mass normalized XLFs for
three bins of $\langle t_{\rm age} \rangle =$~0--2.5, 2.5--6, and
6--13.6~Gyr, which we hereafter refer to as ``young,'' ``middle-age,'' and
``old'' populations, respectively.  We find that the total stellar mass for all
three populations is dominated by old populations with $\simgt$100~Myr.  This
is not surprising, given the much larger timescales spanned by the older bins
(i.e., 0.1--1, 1--5, and 5--13.6~Gyr) compared to the younger bins (i.e.,
0--10~Myr and 10--100~Myr).  The XLFs progress from a shallow-sloped power-law
shape for the young population, due to the stronger $\simlt$1~Gyr population
contribution, to increasingly steeper-sloped broken power-law shapes, and
declining normalization, as the population becomes more dominated by the older
population and older LMXBs.  In Figure~13, We display our best-fit SFH XLFs
({\it black curves\/}) and the contributions from each stellar-age bin ({\it
colored-curves\/}), with the colors of each curve corresponding to the colors
of the SFH bins shown in the top panels.  These curves were constructed by
taking each of the five components in the XLF model shown in Figure~12$a$,
multiplying them by their corresponding stellar mass from the top panels of
Figure~13, and finally dividing these contributions by the {\it total} stellar
mass integrated over the entire SFH.  The sum of all five contributing curves,
plus the CXB contribution, provides the total model ({\it black curves\/} in
Fig~13; see Eqn.~(9) for details).

We note that our model is simple, and other model choices may be more
appropriate.  We experimented with different functional form choices for $K(t)$
and $\alpha_2(t)$ and $\alpha_3(t)$, but did not find material differences or
improvements in the quality of the fits.  Ideally, we would be able to measure
the XRB XLF shapes and normalizations for each of the five SFH bins
independently, without a functional form involving age explicitly; however,
such a model would require at least 10 free parameters, even if we chose to
model the XLFs as single-slope power-laws (e.g., normalization and single
power-law slopes for each of the five SFH bins).  Such a model could certainly
fit the data in M51 well, but the parameter values would not be well constrained.
In future studies that include additional data from other galaxies, we will
re-visit such a procedure (see below for further discussion).  

%
%
\begin{figure}
\figurenum{14}
\centerline{
\includegraphics[width=8cm]{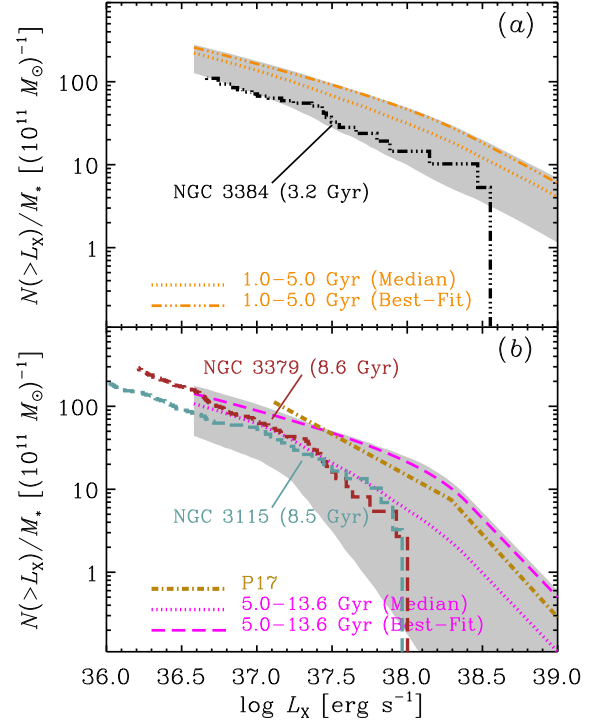}
}
\caption{
($a$) Expanded view of the 1--5~Gyr SFH XLF model.  We show both the best-fit
model ({\it orange triple-dot-dashed curve\/}), as well as the median model
({\it orange dotted curve\/}) with 1$\sigma$ error envelope ({\it gray shaded
region\/}).  For comparison, we show the XLF of field LMXBs in NGC~3384 ({\it
black triple-dot-dashed curve\/}), an elliptical galaxy with mass-weighted
stellar age of $\approx$3.2~Gyr.  ($b$) Expanded view of the 5--13.6~Gyr SFH
XLF model.  Here the best-fit model ({\it dashed magenta curve\/}) lies outside
of the 1$\sigma$ error envelope.  The XLFs for field LMXBs in NGC~3379 and
3115, which have light-weighted stellar ages of 8.6 and 8.5~Gyr, respectively,
have been shown for comparison (Lehmer \etal\ 2014).  Also, the average field
LMXB XLF for nine nearby ellipticals with 9--11.5~Gyr old populations is
plotted as a dotted gold curve (Peacock \& Zepf~2016; Peacock \etal\ 2017).
The comparison field LMXB XLFs appear to show reasonable agreement with the
model XLFs derived here from M51, but show some minor differences in the XLF
shapes.
}
\end{figure}

%
\section{Discussion}
%

In the previous section ($\S$3.4), we presented a comprehensive model for how
the generalized XRB XLF evolves with age, as derived from M51 alone.  We note
that such a model, by definition, is meant to be applicable to describing how
XRB populations form over time in galaxies generally (however, see discussion
of caveats below).  As such, we can make comparisons between our model
estimates and a number of other observations and XRB population synthesis model
predictions.

In Figure~14, we show expanded views of our XLF models for 1--5~Gyr and
5--13.6~Gyr ages, and now include the median model ({\it dotted curves\/}) and
1$\sigma$ error envelopes (i.e., the 16--84\% confidence range).  Note that the
best-fit model for the 5--13.6~Gyr age range (i.e., Fig.~14$b$) mainly lies
outside of the 1$\sigma$ range around the median.  This is driven by the very
long tail in the distribution of the $\alpha_2({\rm 10~Gyr})$ probablity
distribution (see Fig.~11), which leads to large uncertainties in the XRB XLF
at large ages.  For direct comparison, we show the field LMXB XLFs for
elliptical galaxies NGC~3384, NGC~3379, and NGC~3115 (based on Lehmer \etal\
2014), which have light-weighted stellar ages of $\approx$3.2, 8.6, and
8.5~Gyr, respectively.  We also show, in Figure~14$b$, the best-fit average
field LMXB XLF from Peacock \etal\ (2017), which was derived using nine nearby
elliptical galaxies with light-weighted stellar ages spanning
$\approx$9--11.5~Gyr (Peacock \& Zepf~2016).\footnote{We note that the Peacock
\etal\ (2017) field LMXB XLF was provided as the $K$-band luminosity normalized
XLF in the \hbox{0.5--7~keV} band.  We corrected the XLF to our adopted units
by assuming a mean $M_\star/L_K \approx 0.8$~\msol/$L_{K, \odot}$ and a
bandpass correction of $L_{\rm 2-10~keV} = 1.28 L_{\rm 0.5-7~keV}$, which is
appropriate for a power-law \xray\ SED with $\Gamma = 1.7$.}  We note that at a
very peripheral level, the XLFs of the local ellipticals with $\simgt$8~Gyr
populations played a role in our chosen fixed value of $\alpha_3({\rm 10~Gyr})$
(see $\S$3.4 for details), but otherwise played no role in the development of
our SFH XLF model.  Nonetheless, our best-fit models from M51 provide very
reasonable descriptions of the XLFs for all three elliptical galaxies and the
Peacock \etal\ (2017) average field LMXB XLF, but with large uncertainties.

%
%
\begin{figure*}
\figurenum{15}
\centerline{
\includegraphics[width=18cm]{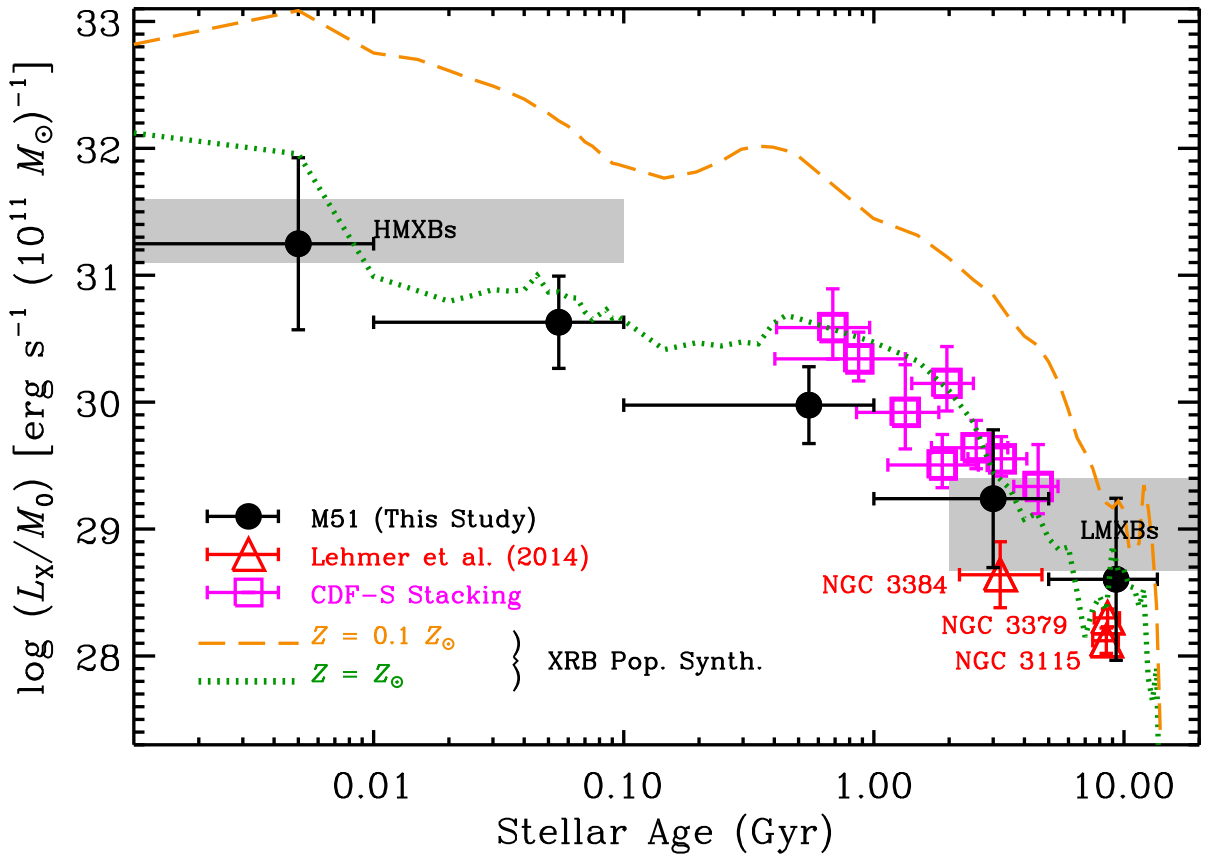}
}
\caption{
Constraints on the \xray\ power output evolution for XRB populations,
normalized by a birth mass of $M_0 = 10^{11}$~\msol.  The black filled circles
represent the XRB \xray\ luminosity evolution as derived from our best-fit SFH
XLF model for M51 (see Fig.~10$a$ and Equation~(13)).  Estimates from HMXB and
LMXB scaling relations ({\it gray rectangles\/}), field LMXBs in three
elliptical galaxies ({\it red triangles\/}), and results from \xray\ stacking
analyses in the $\approx$6~Ms CDF-S ({\it magenta squares\/}) have been plotted
for comparison (see $\S$4 for details).  XRB population synthesis trajectories
have been overlayed for XRB populations born with solar ($Z_\odot$; {\it green
dotted curve\/}) and tenth solar (0.1~$Z_\odot$; {\it orange dashed curve\/})
metallicity.  We find that our constraints from M51 suggest lower XRB \xray\
power output at $\approx$0.6--5~Gyr compared to the CDF-S data; however, this
offset may be related to the significantly higher metallicity in M51
($\approx$2--3~$Z_\odot$) versus the CDF-S galaxies
($\approx$0.8--1.2~$Z_\odot$).
}
\end{figure*}

The direct comparison of our SFH XLF models with observed XLFs from field LMXBs
in other elliptical galaxies of varying ages may not be completely appropriate.
For instance, stellar ages of elliptical galaxies are inferred from optical
spectra, taking advantage of absorption feature strengths and single stellar
population synthesis modeling (see, e.g., McDermid \etal\ 2006;
S{\'a}nchez-Bl{\'a}zquez et al. 2006; Thomas \etal\ 2010).  As such, the ages
are stellar-light weighted, when almost certainly some SFH needs to be
accounted for.  Also, the XRB population modeling in our work is statistical in
nature and does not directly distinguish between LMXB populations in globular
clusters (GCs) versus those found in the field.  The elliptical galaxy
comparison XLFs presented here have GC LMXB populations removed (see Lehmer
\etal\ 2014 for details), since elliptical galaxies are generally much more
rich in GC LMXBs than spiral galaxies like M51, which are dominated by field
LMXBs.  However, M51 may still have a significant GC LMXB population, which we
have not accounted for here.  Despite these issues, our simple age-dependent
SFH XLF model of M51 provides a very similar resemblance to the elliptical
galaxy field LMXB XLFs, albeit with large uncertainties.

By integrating Equation~(10), we can obtain a direct prediction for how the
\xray\ power output of the cumulative XRB population evolves over time
following a star formation event.  Specifically, the \xray\ power output per
unit {\it birth stellar mass}, $M_0$, can be obtained following:
\begin{equation}
L_{\rm X}(t)/M_0 =  \left( \frac{M_\star(t)}{M_0} \right) \int_{L_{\rm
lo}}^{L_c} L_{\rm X} h(L_{\rm X},t) d L_{\rm X},
\end{equation}
where $L_{\rm lo} = 0.01$ (i.e., 10$^{36}$~\lum) is a lower integration limit,
below which the XLF is observed to turn over and contribution from XRBs are
negligible (see, e.g., studies of the XLF in the MW and Magellanic Clouds from
Grimm \etal\ 2002).  The term $M_\star(t)/M_0$ quantifies the present-day
stellar mass to birth mass ratio as a function of age, and is determined by our
stellar population synthesis modeling; $\log M_\star(t)/M_0 =$~$-0.02$,
$-0.10$, $-0.19$, $-0.27$, and $-0.33$ at 5~Myr, 55~Myr, 550~Myr, 3~Gyr, and
9~Gyr, respectively.  In Figure~15, we show $\log L_{\rm X}(t)/M_0$ versus age,
as derived for the M51 population.  Our results suggest that the \xray\ power
output of XRB populations declines by $\approx$3--4 orders of magnitude from
$\approx$10~Myr to $\approx$10~Gyr due to the combined decline in XLF
normalization and steepening in slope with increasing age (see Fig.~12$a$). 

We can compare the trend observed in Figure~15 with expectations estimated from
\xray\ scaling relations.  For instance, at the young stellar-age end, we can
use the $L_{\rm X}$(HMXB)/SFR relation to provide an order-of-magnitude
estimate of $L_{\rm X}/M_0$ for the $\simlt$100~Myr population.  Typically, SFR
values are estimated as the mean SFR over the last 100~Myr, due to the fact
that SFR tracers provide emission on those time scales.  If the \xray\ emission
is indeed associated with the $\simlt$100~Myr old population, then we can make
the following estimate:
$$\langle L_{\rm X}(\simlt {\rm 100~Myr})/M_0 \rangle \approx \left(
\frac{L_{\rm X}{\rm (HMXB)}}{\rm SFR} \right) \left( \frac{1}{100 \times
10^6~{\rm yr}} \right). $$
Recent scaling relation studies have estimated $\log L_{\rm
X}$(HMXB)/SFR~$=$~39.1--39.6 (Lehmer \etal\ 2010, 2016; Mineo \etal\ 2012) or
$\log \langle L_{\rm X}(\simlt {\rm 100~Myr})/M_0 \rangle \approx$~\hbox{31.1--31.6}.
For LMXBs, scaling relations based on both star-forming and elliptical galaxy
populations indicate $\log L_{\rm X}{\rm (LMXB)}/M_\star$~=~29.0--29.6 (e.g.,
Lehmer \etal\ 2010, 2016; Boroson \etal\ 2011; Zhang \etal\ 2012).  These
galaxy samples span effective ages of \hbox{$\approx$2--15~Gyr}, so correcting
to their birth stellar mass implies $\log L_{\rm X}{\rm
(LMXB)}/M_0$~=~\hbox{28.6--29.4} (i.e., $\log M_\star/M_0 =$~$-0.33$ to
$-0.2$).  We note that the quoted scaling relations were derived {\it without}
making any corrections for GC LMXBs, which will enhance the LMXB emission over
M51 for the case of the ellipticals.  In Figure~15, we indicate the estimated
regions for HMXB and LMXB populations based on scaling relations as gray
rectangles with annotations.  We find good agreement, within the uncertainties,
between the HMXB scaling relations and our M51-based predictions for HMXB
emission, while the LMXB scaling relations predict $L_{\rm X}/M_0$ values that
are somewhat higher than our estimates for M51 at $\approx$10~Gyr, albeit still
within errors.  Such a difference, however, could be due to the
disproportionate boost to the scaling relations from the GC LMXB population
(see below).

We can make further comparisons using additional observations that constrain
the XRB \xray\ emission at different ages.  For example, in Figure~15, we show
$L_{\rm X}/M_0$ values for the field LMXBs ({\it red triangles\/}) in NGC~3384,
NGC~3379, and NGC~3115.  As with the XLF models, the integrated $L_{\rm X}/M_0$
values are in good agreement with our M51-based prediction.  This provides some
indication that M51 has a weak contribution from GC LMXBs; however, direct
counterpart studies are required to confirm this.  

An additional constraint on $L_{\rm X}/M_0$ for various age ranges comes from
the stacking analyses of distant galaxy populations from Lehmer \etal\ (2016),
which are based on a $\approx$6~Ms exposure of the \chandra\ Deep Field-South
(CDF-S; Luo \etal\ 2017) and provide measurements of how the $L_{\rm X}$(HMXB)/SFR and $L_{\rm
X}$(LMXB)/$M_\star$ scaling relations evolved since $z \approx 2.3$.  Here, we
utilize the redshift-dependent $L_{\rm X}$(LMXB)/$M_\star$ scaling relation
constraints from Lehmer \etal\ (2016), and estimate mass-weighted stellar ages
for the redshift bins.  These mass-weighted stellar ages were calculated by
first extracting synthesized galaxy catalogs from the {\ttfamily Millenium~II}
cosmological simulation from Guo \etal\ (2011) that had the same SFR and
$M_\star$ selection ranges as those adopted by Lehmer \etal\ (2016).  These
galaxy catalogs contain estimates of the mass-weighted stellar ages for each
galaxy.  The mass-weighted stellar age of the entire galaxy population
(catalog) is then estimated using Equation~(8), and a standard deviation of the
population is calculated to estimate the error.  All values are corrected by
the single stellar population synthesis derived factor of $M_\star(t)/M_0$ to
convert $L_{\rm X}$(LMXB)/$M_\star$ to $L_{\rm X}/M_0$.  

The magenta squares in Figure~15 show the Lehmer \etal\ (2016) estimates of
$L_{\rm X}/M_0$.  The ages evaluated by the CDF-S stacking analyses span
$\approx$600~Myr to $\approx$5~Gyr, corresponding to the $z \approx 2.3$ and $z
\approx 0.4$ stacked populations, respectively.  Over this age range, the
$\approx$6~Ms CDF-S stacked constraints appear to be somewhat elevated above our
M51-based estimates by a factor of $\approx$2.  The derived scaling relations
for galaxy samples in the CDF-S are expected to be appropriate for
representative galaxies (in terms of representing the majority of the stellar
mass) in the Universe, which are dominated by star-forming galaxies on the
``main sequence'' (e.g., Elbaz \etal\ 2007; Noeske \etal\ 2007; Karim \etal\
2011; Whitaker \etal\ 2014).  These galaxies will contain a broad range of
stellar populations, and complex SFHs (like that of M51).  Therefore taking the
globally averaged \xray\ luminosity per unit mass and associating it with a
single mean stellar age for the population may not provide an entirely
equivalent comparison to our M51 results, which have $L_{\rm X}/M_0$ values
based on decomposition of all stellar age contributions.  We can perform the
equivalent operation for M51, however, by computing the galaxy-wide mean
stellar age and extracting the LMXB emission per unit mass.  Doing so, reveals
$\langle t_{\rm age} \rangle \approx 5.2$~Gyr and $L_{\rm X}({\rm LMXB})/M_0
\approx 29.1$ (based on the LMXB XLF derived in $\S$3.2), which is in nearly
perfect agreement with the $L_{\rm X}/M_0$ estimates for the 1--5~Gyr age bin.
It is therefore unlikely that the CDF-S estimates are systematically elevated
from our M51 estimates simply due to differences in how the points were
derived.

Alternatively, the apparent discrepancy between our results for M51 and those
from the $\approx$6~Ms CDF-S stacking analyses may arise due to the high
metallicity of M51 ($Z \approx$~\hbox{1.5--2.5}~$Z_\odot$; e.g., Moustakas
\etal\ 2010) relative to typical galaxies at $z \approx$~0.3--2.3 ($Z
\approx$~\hbox{0.8--1.2}~$Z_\odot$; e.g., Madau \& Fragos 2016).  Population
synthesis predictions have shown, and several studies now seem to confirm, that
the XRB power output per stellar mass declines with increasing metallicity
(e.g., Fragos \etal\ 2013; Basu-Zych \etal\ 2013a,b, 2016; Prestwich \etal\
2013; Douna \etal\ 2015; Brorby \etal\ 2014, 2016; Lehmer \etal\ 2016).  To
clarify the expected level of this metallicity effect, we show in Figure~15 the
Fragos \etal\ (2013) XRB population synthesis model predictions for the
$Z_\odot$ and 0.1~$Z_\odot$ cases.  We find that the CDF-S stacked data at
$\approx$0.6--5~Gyr are in very good agreement with the $Z_\odot$ XRB
population synthesis predictions.  The 0.1~$Z_\odot$ XRB population synthesis
model is almost uniformly an order of magnitude above the $Z_\odot$ case.
Unfortunately, we do not have available XRB population synthesis models
appropriate for 1.5--2.5~$Z_\odot$; however, if the nearly linear trend of
declining $L_{\rm X}/M_0$ with metallicity were to continue, we might expect
that most of the points in M51 would be factors of $\approx$2--3 times lower
(i.e., $\approx$0.3--0.5~dex), and may explain the offset between the M51
and $\approx$6~Ms CDF-S data points at $\approx$0.6--5~Gyr.  

%
\section{Summary and Future Direction}
%

We have presented a new technique for determining the XLF evolution of XRB
populations in nearby galaxies that have resolved XRB populations (e.g., via
\chandra\ observations) and multiwavelength data sufficient for determining
accurate SFHs on subgalactic scales.  We have performed SED fitting of
far-UV--to--far-IR data to construct $M_\star$, SFR, and SFH maps, and we
utilize the $\approx$850~ks cumulative \chandra\ exposure of M51 to constrain
the XRB population demographics within the galaxy.  We spatially segregate
\xray\ source populations within regions of varying sSFR and mean mass-weighted
stellar age, and then self-consistently model how the XLFs vary accross these
regions.  Below, we summarize our key findings.

\begin{itemize}

\item By dividing the galaxy into regions with varying sSFR, we are able to
decompose the XRB XLF into LMXB and HMXB contributions that scale with
$M_\star$ and SFR, respectively ($\S$3.3).  Our results are broadly consistent
with past studies of the scaling of XRB XLFs from actively star-forming galaxy
samples (e.g., Mineo \etal\ 2012) and passive ellipticals (e.g., Zhang \etal\
2012).  However, we find that our inferred LMXB XLF has an excess of $L_{\rm X}
\ge 10^{38}$~\lum\ sources compared to elliptical galaxies (Fig.~8).  This
result is potentially due to the LMXB population in M51 being younger than
typical ellipticals (i.e., $\approx$5~Gyr for M51 and $\approx$10~Gyr for
ellipticals), as was hypothesized in Kuntz \etal\ (2016).

\item When dividing the galaxy into regions based on the local mean stellar
age, we were able to self-consistently model the XRB XLFs using an
age-dependent model where the XLF shape and normalization evolve with time
($\S$3.4).  Our best-fit model indicates that the normalization of the XRB XLF
declines by $\sim$3--3.5 orders of magnitude from $\approx$10~Myr to
$\approx$10~Gyr, while the overall XLF slope steepens over this time period
(Fig.~12).

\item Through a statistical comparison of models, we find that our generalized
evolving XRB XLF model provides a better fit to the data in all subregions of
M51 compared to the LMXB and HMXB decomposition model ($\S$3.4).  In principle,
this model is robust and applicable to XRB populations in other galaxies,
provided the metallicities are similar and the XRBs are associated with the
evolution of the underlying stellar populations.  We find that our XRB XLF
models for the $\approx$3--11.5~Gyr timescale provide good agreement with
observed field LMXBs in elliptical galaxies of comparable ages, providing
independent support for our model predictions ($\S$4 and Fig.~14).

\item By integrating our evolving XRB XLF model with respect to $L_{\rm X}$, we
can predict the total XRB \xray\ power output evolution with age.  These
predictions are in good agreement with those provided by XRB population
synthesis models, high-redshift stacking results, estimates from scaling
relations, and field LMXBs in nearby elliptical galaxies (see $\S$4 and
Fig.~15).

\end{itemize}

The above conclusions provide a step forward in empirically calibrating how XRB
XLFs evolve with age, generally.  However, we expect that the XRB evolutionary
history will also be dependent on the metallicity history, since metallicity is
expected to be a major factor in the formation of XRBs relative to the stellar
population (see Fig.~15 and discussion in $\S$4).  In the near future, we will
apply the techniques developed here to a larger suite of $\sim$20 nearly
face-on spiral galaxies, for which SFHs can be calculated well.  With this
larger sample, we will further expand our analyses to include XRB XLF evolution
for samples separated into metallicity bins.  Our ultimate goal will be to
develop a full suite of age and metallicity dependent XRB XLF models that
self-consistently describe well observed XRB XLFs in all nearby galaxies, but
also \xray\ scaling relations and their redshift evolution.

\acknowledgements

We thank the anonymous referee for their helpful comments, which have improved
the quality of this work.  We gratefully acknowledge support from NASA/ADAP
grant NNX13AI48G (B.D.L., R.T.E., A.Z.).  A.Z. acknowledges funding from the
European Union's Seventh Framework Programme (FP/2007--2013)/ERC Grant
Agreement n.~617001.  We thank Kip Kuntz for discussions on X-ray point-source
cataloging efforts, which have been helpful to the quality of our analyses.

%

%

\end{document}